\documentclass[preprint]{aastex}
\usepackage{natbib}
\newcommand{\lel}{L_{e,1.4}}
\newcommand{\dlogL}{d\log_{10}L}
\newcommand{\msun}{M_{\odot}}

\newcommand{\zmin}{z_{min}}
\newcommand{\zmax}{z_{max}}

\newcommand{\unit}[1]{\ifmmode\,{\rm #1}\else$\,{\rm #1}$\fi}
\newcommand{\wmsh}{\unit{Wm^{-2}ster^{-1}Hz^{-1}}}
\newcommand{\whzmpc}{\unit{W Hz^{-1} Mpc^{-3}}}
\newcommand{\whz}{\unit{W Hz^{-1}}}
\newcommand{\myrmpc}{\unit{\msun yr^{-1} Mpc^{-3}}}
\newcommand{\mum}{\unit{\mu m}}
\newcommand{\mujy}{\unit{\mu Jy}}
\newcommand{\kmsmpc}{\unit{km\,s^{-1}Mpc^{-1}}} 

\newcommand{\ie}{{\it i.e.}}
\newcommand{\eg}{{\it e.g.}}
\newcommand{\etal}{~et~al.\ }

\newcommand{\ho}{$H_0$}
\newcommand{\Om}{$\Omega_m$}
\newcommand{\Ol}{$\Omega_\Lambda$}

\begin{document}

\title{
Faint Radio Sources and Star Formation History 
}
\author{D. B. Haarsma}
\affil{Calvin College, 3201 Burton Street SE, Grand Rapids, MI 49546 {\tt dhaarsma@calvin.edu}}
\author{R. B. Partridge}
\affil{Haverford College, Haverford, PA 19041 {\tt bpartrid@haverford.edu}}
\author{R. A. Windhorst and E. A. Richards\altaffilmark{1}}
\affil{Department of Physics \& Astronomy, Arizona State University, 
Tempe, AZ, 85287-1504 
{\tt rogier.windhorst@asu.edu}, {\tt eric.richards@asu.edu}}
\altaffiltext{1}{Hubble Fellow}

\begin{abstract}

The centimeter-wave luminosity of local radio galaxies correlates well
with their star formation rate.  We extend this correlation to surveys
of high-redshift radio sources to estimate the global star formation
history.  The star formation rate found from radio observations needs
no correction for dust obscuration, unlike the values calculated from
optical and ultraviolet data.  Three deep radio surveys have provided
catalogs of sources with nearly complete optical identifications and
nearly 60\% complete spectroscopic redshifts: the Hubble Deep Field
and Flanking Fields at 12h+62d, the SSA13 field at 13h+42d, and the
V15 field at 14h+52d.  We use the redshift distribution of these radio
sources to constrain the evolution of their luminosity function.  The
epoch dependent luminosity function is then used to estimate the
evolving global star formation density.  At redshifts less than one,
our calculated star formation rates are significantly larger than even
the dust-corrected optically-selected star formation rates; however,
we confirm the rapid rise from $z=0$ to $z=1$ seen in those surveys.

\end{abstract}


\section{Introduction}
\label{intro}

\begin{figure}
\epsscale{0.7}
\plotone{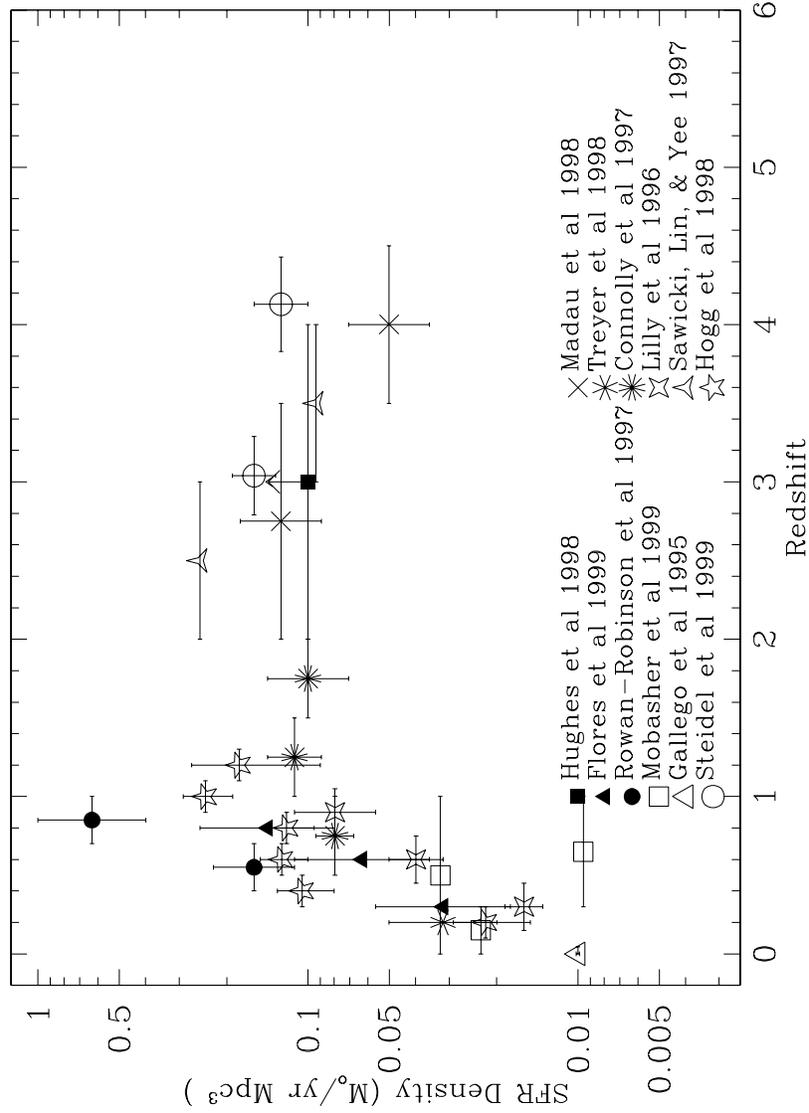}
\caption{Some star formation histories at various wavelengths. All the
data points are scaled to $H_0=50, \Omega_m=1, \Omega_\Lambda=0$, and
a Salpeter Initial Mass Function from 0.1-100$\msun$ (following the
scaling by Baugh\etal1998, their figure 16).  Corrections for dust
extinction calculated by Steidel\etal(1999, their figure 9) were used.
}
\label{fig.sfhist.all-radio}
\end{figure}

In the last few years, a variety of observational methods have been
used to study the global star formation history of the universe at a
range of redshifts.  Figure~\ref{fig.sfhist.all-radio} compiles the
results from several of these studies, scaling all to the same
cosmology and IMF.  In plotting the points, we used the corrections
for extinction by dust calculated by Steidel\etal\nocite{steidel99a}
(1999, their figure 9).  The diagram shows significant scatter in the
star formation density at each redshift.  Most studies agree, however,
that the star formation density rises rapidly from $z=0$ to $z=1$.
Beyond a redshift of 1 it is unclear whether the star formation
density decreases significantly (as suggested at first by
\citealp{madau96a}) or stays roughly constant (for example,
\citealp{steidel99a}).

\nocite{hughes98a}
\nocite{flores99a}
\nocite{rowan-robinson97a}
\nocite{mobasher99a}
\nocite{gallego95a}
\nocite{steidel99a}
\nocite{madau98a}
\nocite{treyer98a}
\nocite{connolly97a}
\nocite{lilly96a}
\nocite{sawicki97a}
\nocite{hogg98a}
\nocite{baugh98a}

Radio observations have important advantages in determining the global
star formation history, and are a useful complement to studies at
other wavelengths.  Unlike calculations based on ultraviolet and
optical observations, there is no need to make uncertain corrections
for dust extinction since the radio emission at $\nu\gtrsim1$~GHz
passes freely through dust.  Compared to far-infrared and
submillimeter studies, interferometric radio observations typically
have better positional accuracy, allowing for more reliable
identifications with objects detected at other wavelengths
\citep{richards99a,downes99a}.  Finally, since far-infrared emission
is due to re-heated dust, the original source of energy (whether star
formation or AGN) can be unclear; radio properties, such as spectral
index and morphology, can help distinguish between these.  We do note,
however, that contamination of the radio flux by emission from AGN is
a problem (addressed in \S\ref{data.limits}).  In addition, relatively
few high redshift star-forming radio sources have as yet been detected
(though current and planned deep radio surveys will rapidly change
that).  As a consequence, the statistical sample used in this work is
small.

Our strategy in this paper is as follows.  We use very sensitive radio
surveys to detect star forming galaxies at high redshift.  Not all of
the sources in these surveys are star-forming (some are probably AGN),
but we deal with this problem by defining data samples that give lower
and upper limits to the star formation history, as described in
\S\ref{data}.  Next, in \S\ref{lumfunc}, we use these data to
determine the evolving luminosity function for star-forming radio
sources.  The redshift and flux distribution of the sources, as well
as the total extragalactic radio background, are used to constrain the
evolution.  In \S\ref{sfhist}, we use the well-known relationship
between radio luminosity and star formation rate (and discuss the
assumption that this relationship holds for all redshifts) to find the
star formation history directly from the observed radio sources.  The
evolving luminosity function is used to correct for faint sources
below the observational detection limits.  We discuss our conclusions
in \S\ref{conclusion}.

Unless stated otherwise, we assume a cosmology of $H_0=50$,
$\Omega_m=1$, and $\Omega_\Lambda=0$, and a non-evolving Salpeter
Initial Mass Function with a stellar mass range of $0.1-100\msun$.  We
use a radio spectral index of $\alpha=0.4$ (where $S\propto
\nu^{-\alpha}$), which is appropriate for faint sources selected at 5
or 8~GHz \citep{windhorst93a,richards00a}.

\section{The Data Sets}
\label{data}

\subsection{Surveys} 
\label{data.surveys}

Three fields have been observed to micro-Jansky sensitivities at
centimeter wavelengths and also have extensive photometric and
spectroscopic data: the Hubble Deep Field (HDF) at 8~GHz
(Table~\ref{tab.hdf}), the SSA13 field at 8~GHz (Table~\ref{tab.13h}),
and the V15 field at 5~GHz (Table~\ref{tab.hdf}).
Table~\ref{tab.data} summarizes information on the three fields, all
of which were observed at the Very Large Array (VLA\footnote{The
National Radio Astronomy Observatory is a facility of the National
Science Foundation operated under cooperative agreement by Associated
Universities, Inc.}).  Tables~\ref{tab.hdf}, \ref{tab.13h}, and
\ref{tab.v15} list the individual sources in each field; the columns
are as follows:
\\{\it Column 1:} The source name. 
\\{\it Columns 2,3:} $S_8$, $S_5$, and $S_{1.4}$ are the radio flux
densities at 8~GHz, 5~GHz, and 1.4~GHz respectively.  If a measured
value of $S_{1.4}$ is not available, we use the spectral index shown
in Column~4 to calculate $S_{1.4}$ and list the value in parentheses
in Column~3.
\\{\it Column 4:} The radio spectral index, $\alpha$, defined as
$S\propto\nu^{-\alpha}$.  If 1.4~GHz measurements are not available,
we use assume a spectral index of 0.4 unless this violates the survey
detection threshold at 1.4~GHz.
\\{\it Column 5:} The primary beam correction factor, $B_i$ (see
\S\ref{data.pbcor}, eq.~\ref{eq.pbcor}).
\\{\it Column 6:} The galaxy type (see \S\ref{data.id}): {\it sim}
refers to spiral, irregular, or merger; {\it el} refers to elliptical;
{\it fr} refers to faint ($I>25$) or red ($I-K>4$); and {\it un}
refers to unknown type or undetected.  The galaxy type in parentheses
is the assumed galaxy type in the case of {\it un}.
\\{\it Column 7:} The redshift type (see \S\ref{data.z}): {\it sp} refers to
spectroscopic, {\it ph} refers to rough photometric (based on I or HK$'$
magnitudes), {\it a} refers to random assignment, and {\it afr} refers
to random assignments for galaxies of type faint/red.
\\{\it Column 8:} The redshift used in calculations.
\\{\it Columns 9,10:} The I and HK$'$ magnitudes.  K magnitudes are
converted to HK$'$ magnitudes by $K = HK' - 0.3$.
\\{\it Column 11:} The maximum redshift, $z_{max}$, at which the
galaxy would have been detected, based on its emitted luminosity
$\lel$.
\\{\it Column 12:} The log of the luminosity emitted at 1.4~GHz.
\\{\it Column 13:} The star formation rate for each individual galaxy,
derived from its radio luminosity ($\lel$) and equation 15.  If the
source is elliptical (or assumed elliptical), the emission is probably
contaminated by AGN and the calculated star formation rate is only an
upper limit.  For some of these sources, the AGN contamination causes
the calculated star formation rate to be unphysically large (greater
than 5000$\msun/yr$), so it does not provide an interesting upper
limit and we do not list it; we do, however, include these sources in
our ``upper'' sample to give conservative upper limits on our results.
\\{\it Column 14:} The samples to which the source was assigned (see
\S\ref{data.limits}): {\it U} refers to the upper sample, {\it M} to the middle
sample, and {\it L} to the lower sample.

For the first time we have a sample of microJy radio sources with
nearly complete optical identifications and nearly 60\% complete
spectroscopic redshift measurements.  Others are doing similar work on
faint star forming radio galaxies
(\citealp{hopkins99a,mobasher99a,benn93a}; \citealp*{gruppioni99b})
with larger catalogs of sources.  While our survey samples have fewer
sources, we have generally more sensitive radio flux limits and more
complete optical followup.  In principle, this allows us to probe
higher redshifts, and to increase the fraction of sources identified
with star-forming galaxies (see \S\ref{data.id}).

\subsection{Primary beam corrections} 
\label{data.pbcor}

In each of the three radio surveys, the flux threshold varies
significantly across the field due to the shape of the primary beam
response of the VLA antennas.  The flux limit listed in
Table~\ref{tab.data} is for the center of the field; the limiting flux
$S_{lim}$ increases to the edge of the field by about 1.5 for SSA13,
and by about 10 for the larger HDF and V15 fields.  To find the total
surface density of sources $n$ (the number of sources per angular area
on sky), we determine the contribution of each source by considering
the portion of the field in which that source could have been
detected.  Therefore, each source $i$ contributes
\begin{equation}
	B_i = \frac{ 1 }{ A_i(S_i,S_{lim}) }
\label{eq.pbcor}
\end{equation}
to the surface density of sources, where $A_i$ is the solid angle on
the sky in which the flux $S_i$ of source $i$ would be greater than
the sensitivity limit of the radio survey $S_{lim}$.  A faint source
which could only be detected at the center of the field (small $A_i$)
contributes more to the average surface density $n$ than a strong
source which could be detected over the entire primary beam area
(large $A_i$).  To determine $A_i$ for each source, we used the shape
of the VLA primary beam.\footnote{The gain of the VLA primary beam is
well-matched by $g(r) = [cos( \frac{ -0.23226 + 74.567639 r
\nu}{57.2957795} )]^6$ where $r$ is the distance from the beam center
in degrees and $\nu$ is the observing frequency in GHz
\citep{oort85a}.}  The total surface density of sources for a survey
is then
\begin{equation}
	n = \Sigma B_i
\end{equation}

Other instrumental effects that affect the point source sensitivity
across the field, such as bandwidth smearing, time delay smearing, and
geometrical smearing \citep{richards00a}, are negligible in the three
VLA surveys.  \citet{pascarelle98a} discuss the importance of surface
brigtness corrections in determining star formation history. However,
the relatively low resolution of these radio surveys (3$''$ for the
HDF and 3-10$''$ for the SSA13 and V15 fields), combined with the
resolution correction as a function of flux density
\citep{windhorst90a,windhorst93a}, suggests that few of these sources
are resolved and thus no correction for surface brightness biases have
been made.

\subsection{Redshifts}
\label{data.z}

The redshifts for the sources in the sample are either spectroscopic
measurements, estimates from I or HK$'$-band magnitudes, or random
assignments (see Tables~\ref{tab.data} through \ref{tab.v15} and
Figure~\ref{fig.nzztype}).  About 58\% of the sources have
spectroscopic redshifts.  The highest spectroscopic redshift in the
sample is a source at $z=4.42$ in the HDF \citep{waddington99a};
however, there is some evidence that this source may contain an AGN,
so that its radio flux is not dominated by star formation (see
\S\ref{data.limits} for how this source is treated in the
calculations).

\begin{figure}
\epsscale{0.8}
\plotone{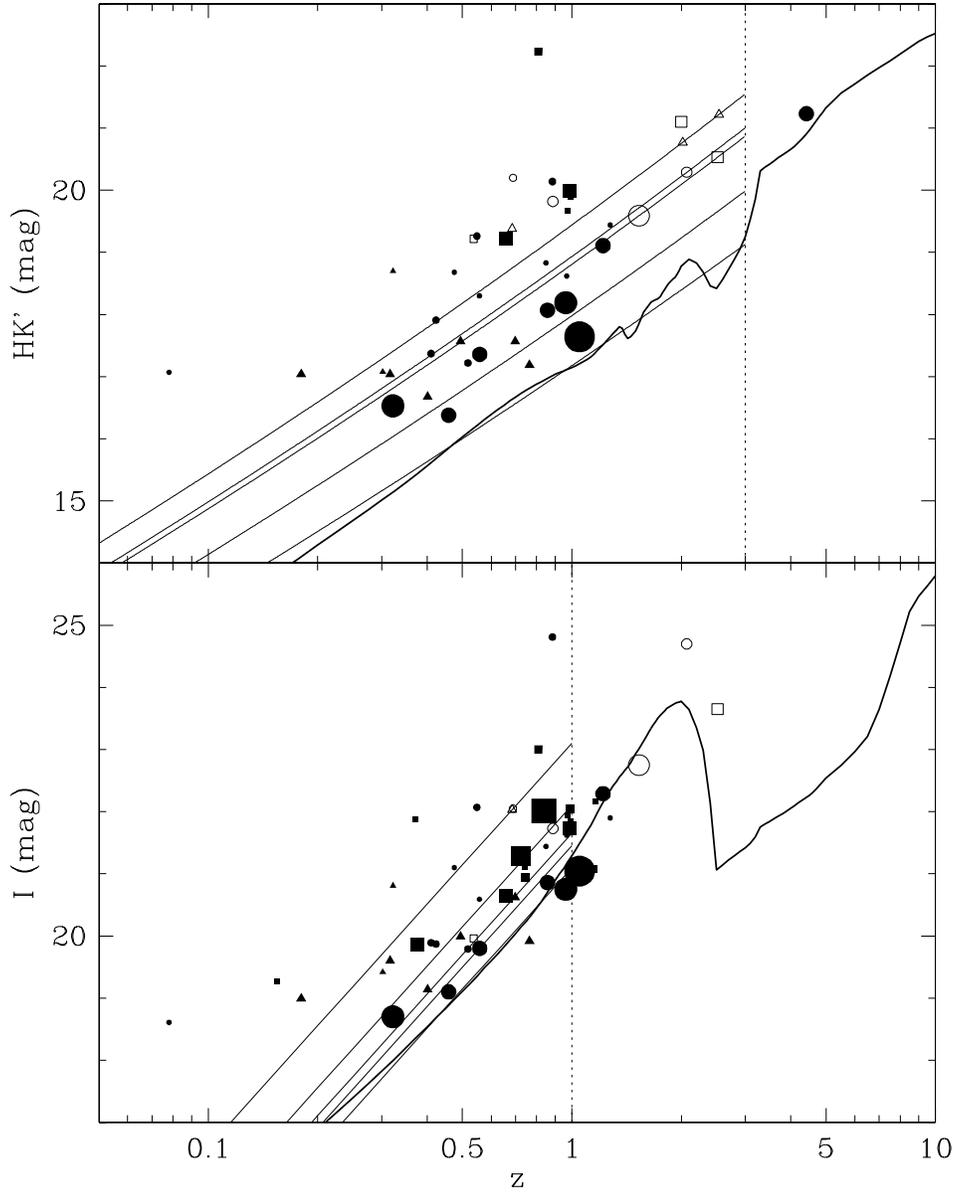}
\caption{Estimation of photometric redshifts (see \S\ref{data.z}).
Circles are sources in the HDF field, triangles in the SSA13 field,
and squares in the V15 field.  The symbol size is proportional to
radio flux density.  Solid symbols are spectroscopic redshifts, hollow
symbols are photometric redshifts.  The thick line is the model for
the $I(z)$ and $HK'(z)$ relationships for faint radio sources
\citep{windhorst94b}.  The thin lines are parallel to the model curve,
but are offset vertically to fit the spectroscopic redshifts in
different radio flux density ranges.  From bottom to top these flux
ranges are: $S_8>300\mujy$, $300\mujy>S_8>100\mujy$,
$100\mujy>S_8>30\mujy$, $30\mujy>S_8>18\mujy$, and $S_8<18\mujy$.  }
\label{fig.IKz}
\end{figure}

\begin{figure}
\epsscale{0.8}
\plotone{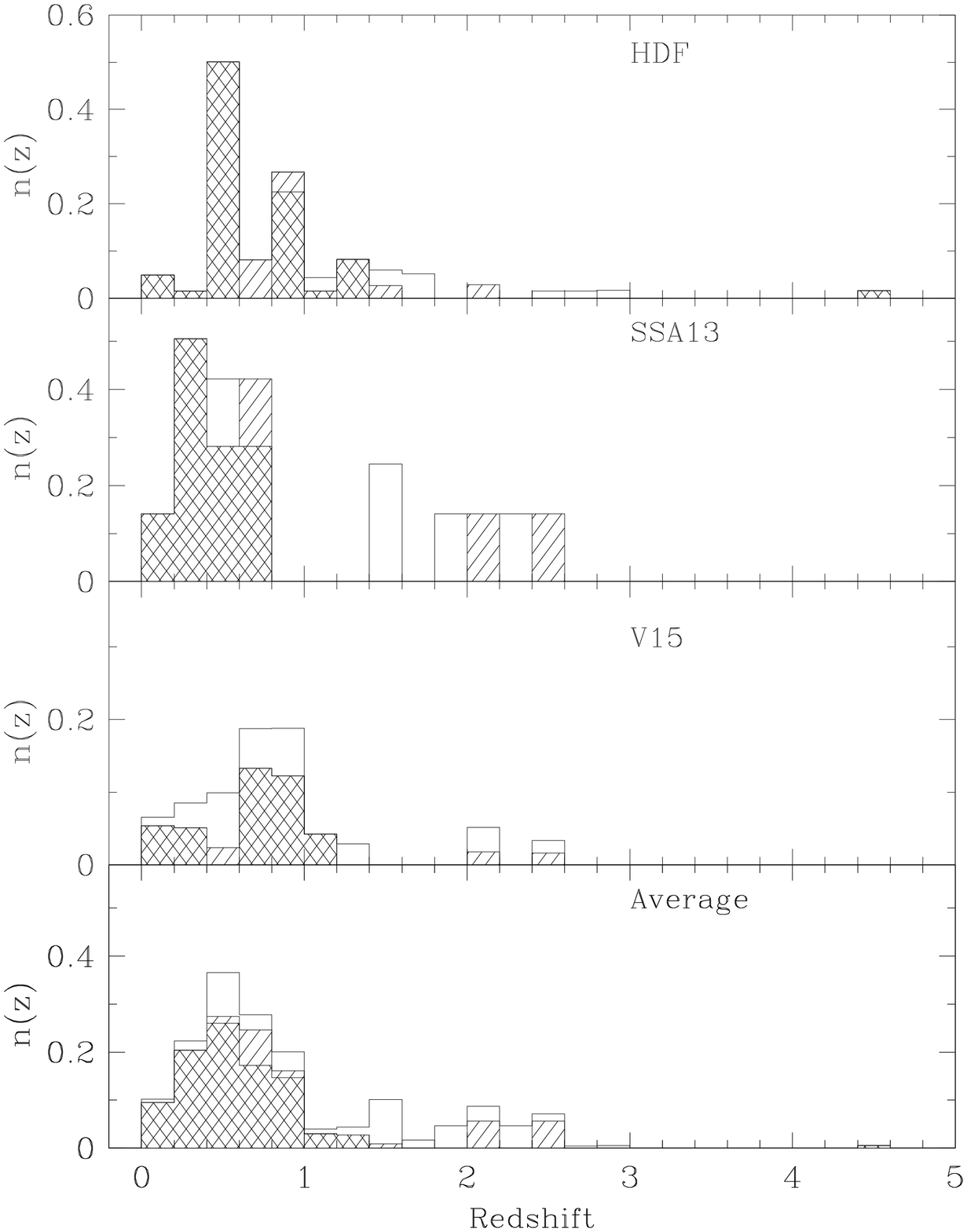}
\caption{ Redshift distribution of sources, in number/arcmin$^2$ and
corrected for the primary beam (\S\ref{data.pbcor}).  Redshifts were
measured spectroscopically {\it (cross-hash)}, estimated from I- or
HK$'$-band magnitudes {\it (hash)}, or assigned {\it (blank}, see
\S\ref{data.z} for how assignments were made).  }
\label{fig.nzztype}
\end{figure}

For 13\% of the sources, approximate redshifts were found from I
and HK$'$ band magnitudes.  \citet{windhorst94b} used Bruzual-Charlot
models (1993)\nocite{bruzual93a} to find the dependence of I and HK$'$
magnitude on redshift for milliJy radio sources; these models are
plotted in Figure~\ref{fig.IKz}.  When comparing these models with the
45 spectroscopic redshifts in our sample of fainter microJy radio
sources, we found a significant dependence on radio flux density.  We
compensated by shifting the magnitude scale of these models to fit the
I or HK$'$ values of our sources that {\it do} have spectroscopic
redshifts, with different shifts for different radio flux ranges (as
listed in the caption of Figure~\ref{fig.IKz}).  We then used the
revised curves to estimate redshifts for our sources that have I or
HK$'$ band magnitudes but not spectroscopic redshifts.  The resulting
redshifts are crude, but are better than random assignments.  We
converted $K$ magnitudes to $HK'$ using $K=HK'-0.3$ \citep{barger99a}.
The $I(z)$ model is double valued for $z\gtrsim1$, and the $HK(z)$
model increases sharply for $z\gtrsim3$, so when the I or HK$'$
magnitude of a source indicated a redshift above these limits, we
randomly assigned a redshift instead (see below).

\epsscale{1.0}

For the remaining 29\% of the sources, neither spectroscopic redshifts
nor rough photometric redshifts were available.  Rather than removing
these sources from the sample, we assigned redshifts in the following
manner.  We first separated the sources into two groups: those with
very faint (I$>25$) or red (I-K$>4$) optical identifications, and
those with brighter optical identifications.  The group with brighter
identifications (10 sources) were assigned redshifts randomly selected
from the list of spectroscopic redshifts of star-forming galaxies in
the sample.  The group with very faint or red optical identifications
(12 sources) are probably star-forming galaxies at redshifts greater
than 1 (\citealp{richards99d}; \citealp*{barger00a}). These sources
were assigned redshifts randomly in the range $z=1-3$.

\epsscale{1.0}

The redshift distributions and total source densities of the three
surveys are strikingly different (see Figure~\ref{fig.nzztype}).  The
HDF and SSA13 surveys were both performed at 8~GHz with similar flux
limits, yet the average source density (including all redshifts) is
1.3 sources/arcmin$^2$ in the HDF, but 2.7 sources/arcmin$^2$ in the
SSA13 field, over twice as great.  The total source density of the V15
field at 5~GHz is 0.7 sources/arcmin$^2$, which is nearly the same as
the HDF field when the differences in flux limit and observing
frequency are taken into account.  The redshift distribution in the
three fields peaks at somewhat different redshifts (see
Figure~\ref{fig.nzztype}), possibly due to galaxy superclustering or
other high-redshift structure, although all three fields peak at $z<1$
and have a long tail which extends to $z\sim3$.  The differences in
the redshift distribution of the fields are probably due to cosmic
variance (note that each field is sampling only a small solid angle).
Since these fields were generally chosen to be free of bright sources,
the number counts may be too low in the HDF and V15 fields rather than
too high in the SSA13 field (indeed, \citealp{richards00a} reports a
deficit of radio sources detected at 1.4 GHz in the HDF).  To deal
with the differences between fields, we average the three data sets
together in our calculations.

\subsection{Optical Identifications}
\label{data.id}

\begin{figure}
\epsscale{0.75}
\plotone{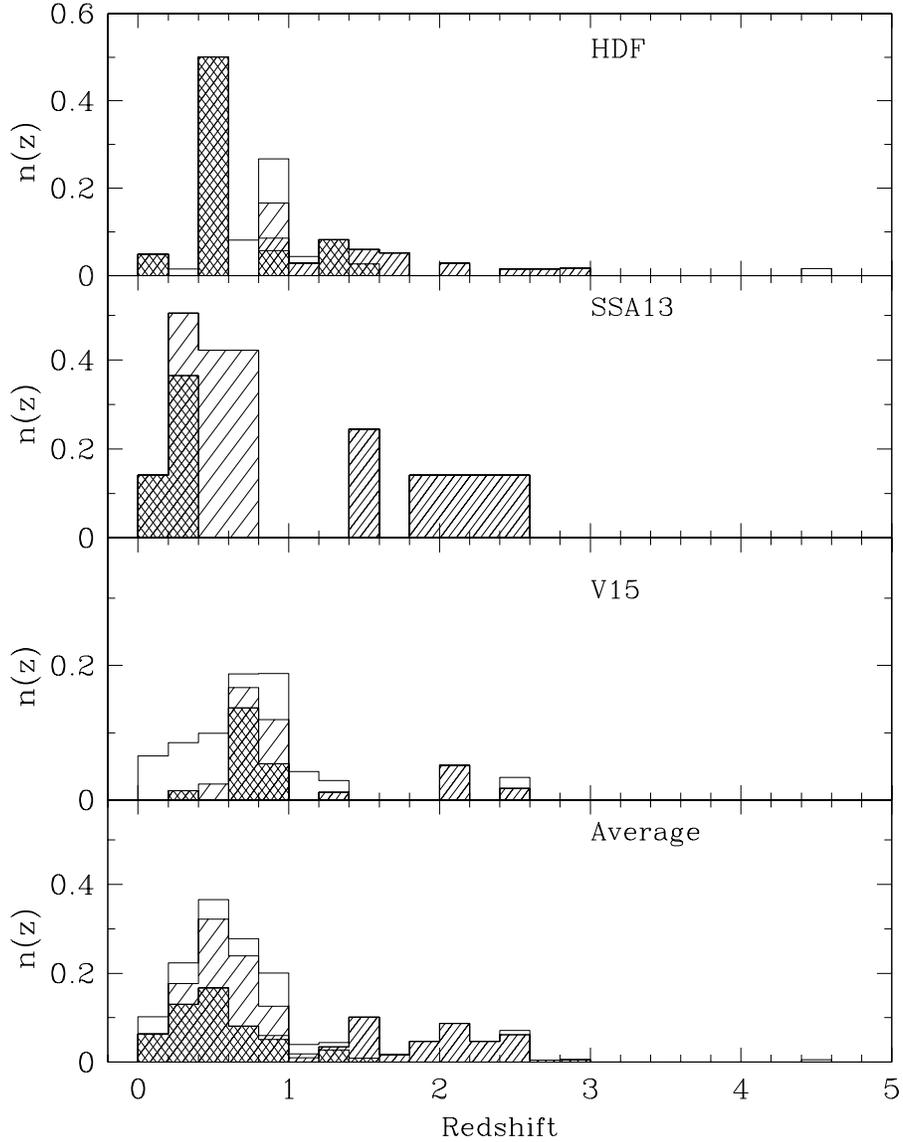}
\caption{ Redshift distribution of sources, {\it separated by galaxy
type}, in number/arcmin$^2$ and corrected for the primary beam shape
(\S\ref{data.pbcor}).  Indicated are spiral, irregular, or merging
galaxies {\it (cross-hash)}, very faint or red optical identifications
{\it (narrow hash)}, unknown or unclear identifications {\it (broad
hash)}, and elliptical or emission line galaxies {\it (blank)}.  Known
QSOs and stars (just four sources in the three surveys) are not
included in the figures or our calculations. }
\label{fig.nzgaltype}
\end{figure}

Figure~\ref{fig.nzgaltype} indicates the nature of the available
optical identifications of the radio sources.  Known QSOs were removed
from the sample (two from the SSA13 field, one from V15), as was one
star in the V15 field and are not shown in the figure.  For the
remaining sources, the three fields appear to be significantly
different.  In the HDF, about 50\% of the radio sources have
star-forming counterparts (spirals, mergers, and irregulars;
\citealp{richards98a}), another 30\% are in the red/optically faint
category discussed above (several of which are identified with bright
sub-mm objects and may include star-forming galaxies;
\citealp{barger00a}), and only 20\% are identified with elliptical
galaxies which presumably are associated with low luminosity AGN.  In
the SSA13 field, 50\% are identified with star-forming or
red/optically faint galaxies, and 50\% are of unknown type.  In the
V15 field, 15\% of the sources have unknown galaxy types, and 40\% of
the sources are likely to be star-forming or in the red/optically
faint population \citep{hammer95a}.  A further 35\% are claimed to be
elliptical/AGN counterparts based on deep I-band images from the
Canada-France Hawaii Telescope, and another 15\% are classified by
Hammer\etal as AGN based on emission line studies.  Thus,
\citet{hammer95a} report a larger fraction of low-luminosity AGN in
the V15 field (50\%) than observed in the HDF (20\%).  However,
HST/WFPC2 images of these identifications from the Groth Strip survey
\citep{hammer96a} show some ambiguity in the optical identifications,
indicating a higher fraction of late type galaxies than originally
reported by \citet{hammer95a}; thus some uncertainty remains.  Also,
the V15 field lies only 20~arcmin from the cluster associated with
3C295, and there is another supercluster or redshift structure at $z =
0.98$ within the field \citep{lefevre94a}.  These structures have
probably caused some bias in the identifications, increasing the
fraction of early type radio galaxies.  Thus we adopt the statistics
from the HDF and SSA13 surveys, which imply that about 50\% of the
sources have disk or late type galaxy counterparts, 30\% have
red/optically faint identifications, and 20\% are associated with
ellipticals and low luminosity FR I type AGN.

\epsscale{1.0}

\subsection{Strategy for dealing with incomplete identifications and redshifts}
\label{data.limits}

\begin{figure}
\epsscale{0.75}
\plotone{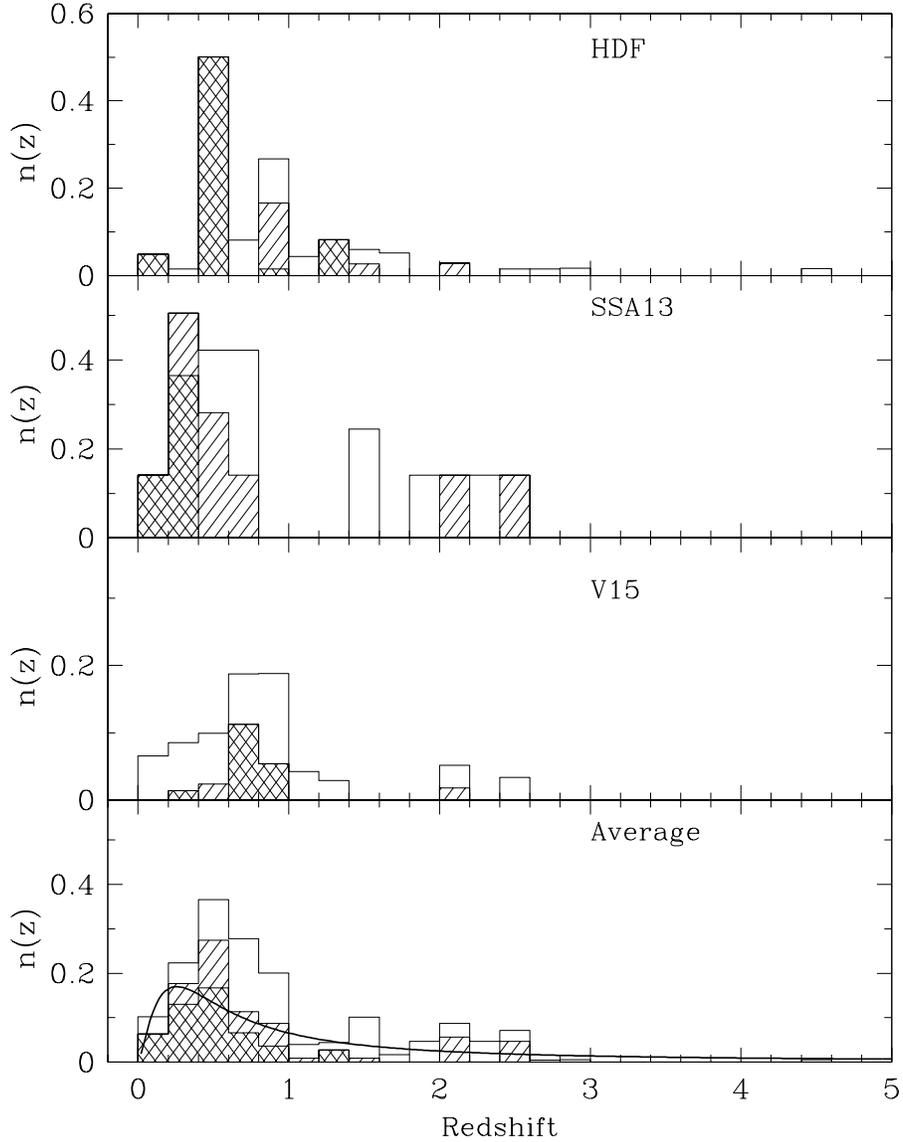}
\caption{ Redshift distribution of sources, separated into
the``lower'' sample {\it (cross-hash)}, ``middle'' sample {\it
(cross-hash and hash)}, and ``upper'' sample {\it (all sources
shown)}.  These samples are used in \S\ref{lumfunc} and \S\ref{sfhist}
to calculate a ``middle'' value with lower and upper limits; see
\S\ref{data.limits} for how samples are defined.  The curve is the
model prediction based on a fit to the middle sample shown here and 
other data (see discussion in \S\ref{lumfunc.modelfit}).  }

\label{fig.nzlimit}
\end{figure}

The goal of this work is to calculate the global star formation
history based on star-forming radio sources.  Since not all of the
sources are star-forming, and only 58\% have spectroscopic redshifts,
we must define the target population carefully.  To do this, we
separate the data into subsets in order to calculate lower and upper
limits on the luminosity function and hence the star formation history
(see Figure~\ref{fig.nzlimit}):

\begin{itemize}

\item a ``lower limit'' sample (23 sources): only those
sources which are {\it both} identified with spiral/irr/mergers {\it
and} have spectroscopic redshifts.  These are the sources which
definitely belong in the population of interest.

\item a ``middle value'' sample (37 sources): only those sources 
for which two criteria are met:
\begin{itemize}
\item redshift is spectroscopic {\em or} based on I- or HK$'$-band magnitude
(no randomly assigned redshifts)
\item galaxy type is spiral/irr/merger, {\em or} faint/red.  In
addition, about 80\% of the ``unknown'' identifications are assumed
to be spiral/irr/merger or faint/red and are included here. The
highest redshift source ($z=4.42$ in the HDF) is not included here
because \citet{waddington99a} argue that it contains an AGN component.
\end{itemize}
{\it This sample is our best estimate of the true redshift
distribution of star-forming radio galaxies.}

\item an ``upper limit'' sample (all 77 sources): all sources
(including identifications with elliptical, emission line, and a few
Seyfert galaxies, but not verified QSOs which were removed from
sample).  Redshifts were assigned for those sources without
spectroscopic or I- or HK$'$- band estimates.  This sample shows the
maximum star formation rate that the data would allow, assuming {\em
all} radio flux from all detected sources is due to star formation and
that the redshift estimates are correct.

\end{itemize}

\epsscale{1.0}

\section{Evolution of the luminosity function}
\label{lumfunc}

Next we determine the evolution of the luminosity function for this
population of faint star-forming radio galaxies, using the data
described in \S\ref{data}.  In \S\ref{sfhist.model} we will use this
evolving luminosity function to build a model of the star formation
history.  Since our data contains very few low-redshift objects, we
can not fit for the shape of the local luminosity function.  Instead,
we use the local luminosity function found by \citet{condon89a}, and
fit for the evolution of that function in luminosity and number
density.  In \S\ref{lumfunc.data} we calculate the luminosity
function directly from the data, in \S\ref{lumfunc.model} we describe
the evolution model, and in \S\ref{lumfunc.modelfit} we describe the
observational constraints on that model and the resulting best fit.

We convert all observed luminosities to a rest-frame frequency of
1.4~GHz, since most of the work on the local luminosity function has
been done at this frequency.  Our samples are defined at 5 and 8~GHz,
but some sources have also been detected at 1.4~GHz.  We use the
observed 1.4~GHz flux densities when available, and for the remaining
sources we assume a spectral index of $\alpha=0.4$ (see \S\ref{intro})
to convert to 1.4~GHz, unless this violates an observed limit on the
1.4~GHz flux density.  All source luminosities are then converted from
the observed 1.4~GHz value to their rest-frame 1.4~GHz value.  Thus,
the observed luminosity of each galaxy $L_{o,\nu}$ at an observing
frequency $\nu$ and redshift $z$, is converted to the emitted
luminosity at 1.4~GHz rest-frame frequency using
\begin{equation}
	\lel = L_{o,\nu} 
	        \left( \frac{\nu }{ 1.4\unit{GHz}} \right)^\alpha 
	        (1+z)^\alpha
\label{eq.LemitLband}
\end{equation}
We define the luminosity function $\phi(\lel)$ as the number per
comoving Mpc$^3$ per $\dlogL$ of star-forming radio sources with
emitted luminosity $\lel$(W/Hz)at 1.4~GHz.

\subsection{Luminosity function estimated from the data}
\label{lumfunc.data}

We can calculate the luminosity function directly from the detected
sources for those luminosity and redshift ranges which are sampled by
the data sets described in \S\ref{data}.  For each bin in luminosity
($L_{min}<\lel<L_{max}$) and redshift ($\zmin<z<\zmax$), the
luminosity function is
\begin{equation}
	\phi(\lel,z)\dlogL = \Sigma_i  \frac{B_i}{V_c[\zmin,\zmax(L_i)]}
\label{eq.lumfunc.data}
\end{equation}
where $B_i$ is the surface density, corrected for the primary beam
(eq.~\ref{eq.pbcor}), and $\zmax(L_i)$ is the largest $z$ in the bin for
which the luminosity of the source $L_i$ was above detection limit of
survey.  $V_c$ is the comoving volume (in Mpc$^3$) between $\zmin$ and
$\zmax$ for solid angle $\Delta\Omega$,
\begin{eqnarray}
	V_c(\zmin, \zmax, \Delta\Omega) 
	& = & \int d\Omega \int r^2(z) dr \nonumber \\
	& = & \frac{\Delta\Omega    }{ \unit{ster}                }
	\left(\frac{\unit{ster}     }{ 1.18\times10^7\unit{arcmin^2}}\right)
	\frac{[r^3(\zmax) - r^3(\zmin)]  }{ 3                    },
\label{eq.covol}
\end{eqnarray}
where the comoving distance is 
\begin{equation}
	r(z) = \frac{2c}{H_0} \left( 1 - \frac{1}{\sqrt{1+z}} \right)
\end{equation}
for our assumed cosmology (see \S\ref{intro}).  

\begin{figure}
\plotone{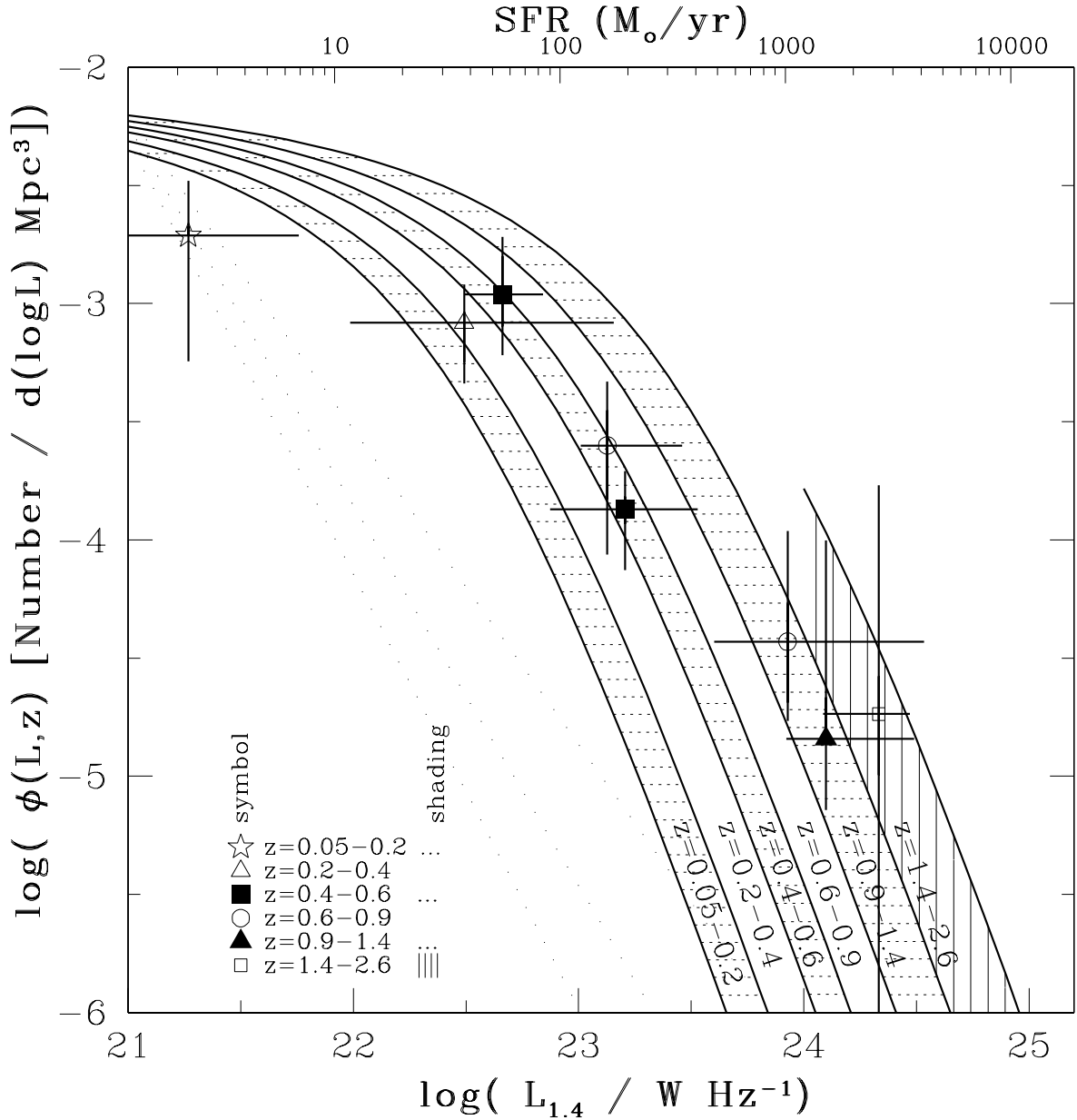}
\caption{ Evolving luminosity function for faint star-forming radio
sources.  Data points are averaged over the three surveys.  Symbol
shapes and shading correspond to the redshift ranges
indicated. Horizontal error bars indicate the range of source
luminosities in the bin.  Vertical error bars are the larger of the
Poisson errors or the lower/upper limits (see \S\ref{lumfunc.data}).
The curves are the model evolving luminosity function, found from a
fit to these and other data (see discussion in
\S\ref{lumfunc.modelfit}). }
\label{fig.lumfunc_evolve}
\end{figure}

The binned luminosity function was calculated using
eq.~\ref{eq.lumfunc.data} for the lower, middle, and upper samples
described in \S\ref{data.limits}, using the average of the three
surveys (bottom panel of Fig.~\ref{fig.nzlimit}).  The result is shown
in Figure~\ref{fig.lumfunc_evolve}, where the data points are from the
``middle'' sample, and are plotted at the average of the luminosities
in each bin.  Vertical error bars are either the lower and upper
limits (from the samples described in \S\ref{data.limits}), or the
Poisson errors ($1/\sqrt{N}$ weights from the number of galaxies per
bin), whichever is larger (generally the Poisson errors dominate for
the low redshift data, and the lower/upper sample limits dominate for
the high redshift data).  Horizontal error bars are the range of
source luminosities in each bin.  Bins were chosen such that each
contains 4 to 6 galaxies (except for the lowest redshift bin which has
only 2 galaxies).

\subsection{Description of evolving luminosity function model}
\label{lumfunc.model}

We now build a model of the evolving luminosity function in order to
compare it to several observables.  In \S\ref{lumfunc.modelfit} we
describe the observational constraints, the fitting process, and
the resulting fit of this model to the observed data.  Here we describe
the model and its free parameters.

We use the local 1.4~GHz luminosity function for star-forming/spiral
galaxies from Condon \nocite{condon89a} (1989, eq. 8 and discussion after 
eq. 7), adopting different notation,
\begin{eqnarray}
\log_{10}[\phi(\lel)] \dlogL = 
	28.83 + Y  - 1.5\log_{10}\lel  \nonumber \\
      - \left[ B^2 + \frac{1}{W^2} (\log_{10}\lel - X)^2 \right]^{1/2} \dlogL,
\label{eq.lumfunc.C89}
\end{eqnarray}
with the fitted parameters for star-forming galaxies of $Y=2.88$,
$X=22.40$, $W=2/3$, and $B=1.5$.  The factor of 28.83 includes unit
conversions and the conversion from magnitudes ($d\log_{2.5}L$) to
base 10 ($\dlogL$).

To describe the evolution of the luminosity function, we use the
functional form suggested by Condon \nocite{condon84a} (1984b,
eq.~24), a power-law in $(1+z)$ with an exponential cut-off at high
redshift.  The luminosity evolves as
\begin{equation}
	f(z) = (1+z)^Q \exp\left[ -\left(\frac{z}{z_q}\right)^q \right],
\label{eq.lumevol}
\end{equation}
and the number density evolves as
\begin{equation}
	g(z) = (1+z)^P \exp\left[ -\left(\frac{z}{z_p}\right)^p \right].
\label{eq.numevol}
\end{equation}
This gives six free parameters $\{Q,q,z_q,P,p,z_p\}$ to use in
describing the evolution.  Thus the general expression for the
evolving luminosity function is \citep{condon84b}
\begin{equation}
	\phi(\lel,z) = g(z) \:\: \phi\!\left( \frac{\lel}{f(z)}, 0 \right).
\label{eq.evolvedlumfunc}
\end{equation}

Once we know the evolving luminosity function, it can be used to
predict the observed redshift distribution, $n(z)$.  The number of
sources between $\zmin$ and $\zmax$ that could be detected in a survey
of angular area $\Delta\Omega$ and flux limit $S_{lim}$ at frequency
$\nu$ is
\begin{equation}
	n(z) = 
	V_c(\zmin,\zmax,\Delta\Omega) \int_{L'(S_{lim},z)}^{\inf} \phi(\lel,z) \; \dlogL
\end{equation}
where the lower limit of the integral is the luminosity corresponding
to the flux limit $S_{lim}$ at the redshift $z$.  The comoving 
volume $V_c$ is defined in eq.~\ref{eq.covol}.

The evolving luminosity function can also be used to predict the
extragalactic background due to this population.  The background
intensity at observing frequency $\nu_0$ is \citep{dwek98a}
\begin{equation}
	I(\nu_0) = \frac{1}{4\pi} \int \rho(\nu,z) 
	           \left| \frac{c dt}{dz} \right| dz 
\label{eq.background}
\end{equation}
where the luminosity density, $\rho$, emitted at redshift $z$ and
frequency $\nu$ is found from the luminosity function,
\begin{equation}
	\rho(\nu, z) = 
	\int \lel \; \phi(\lel,z) \dlogL,
\label{eq.lumdens}
\end{equation}
and
\begin{equation}
	\left| \frac{dt}{dz} \right| = \frac{1}{H_0 (1+z)^{5/2}}
\end{equation}
for the assumed cosmology (\S\ref{intro}).

\subsection{Fitting the evolving luminosity function model to the data}
\label{lumfunc.modelfit}

We can now compare the evolution model to the observed data in order
to fit for the evolution parameters.  The model is constrained by 
three observables:
\begin{enumerate}
\item The redshift distribution, $n(z)$.  We use the ``middle'' sample
(defined in \S\ref{data.limits}) for the average of three surveys,
shown as the hashed area in the bottom panel of Figure~\ref{fig.nzlimit}.
\item The observed luminosity function, shown as the data points in 
Figure~\ref{fig.lumfunc_evolve}.  We use the error bars shown in the
figure (the larger of lower/upper limits and Poisson errors).
\item The extragalactic radio background, which is an important
constraint on the integral of the luminosity function.
\end{enumerate}
The first two constraints are not independent from each other, but both
are needed.  In order for the observed luminosity function to have 4-6
sources per bin, only coarse redshift resolution is possible; the
$n(z)$ distribution allows for more detailed redshift information, but
does not include the luminosity information.

The extragalactic radio background is about half due to star-formation
activity and half due to AGN (see discussion in \citealp{haarsma98a}).
In order to isolate the part of the radio background due to star
formation, we use the far-infrared (FIR) background found by DIRBE of
$1.15\pm0.20\times10^{-20}\wmsh$ near 200\mum\ \citep{hauser98a},
assumed to be due primarily to star formation.  The FIR-radio
correlation (\citealp*{helou85a}; \citealp*{condon91a}) can then be
used to predict the portion of the radio background due to
star-formation, which at 1.4~GHz is $\rho =
3.2\pm0.6\times10^{-23}\wmsh$.

For each trial set of evolution parameters $\{Q, q, z_q, P, p, z_p\}$,
we calculate the model prediction for $n(z)$, the evolving luminosity
function, and the radio background due to star formation. The
evolution parameters are adjusted to improve the model fit to the
three data constraints, using a downhill simplex algorithm
\citep{recipes2} to find the global $\chi^2$ minimum.  Since the
$n(z)$ and luminosity function constraints are not independent from
each other, the reduced $\chi^2$ can not be used to calculate the
``goodness of fit'' or to quantitatively compare the quality of
different fits, but its minimum still indicates the parameters of the
best available fit.

\begin{figure}
\plotone{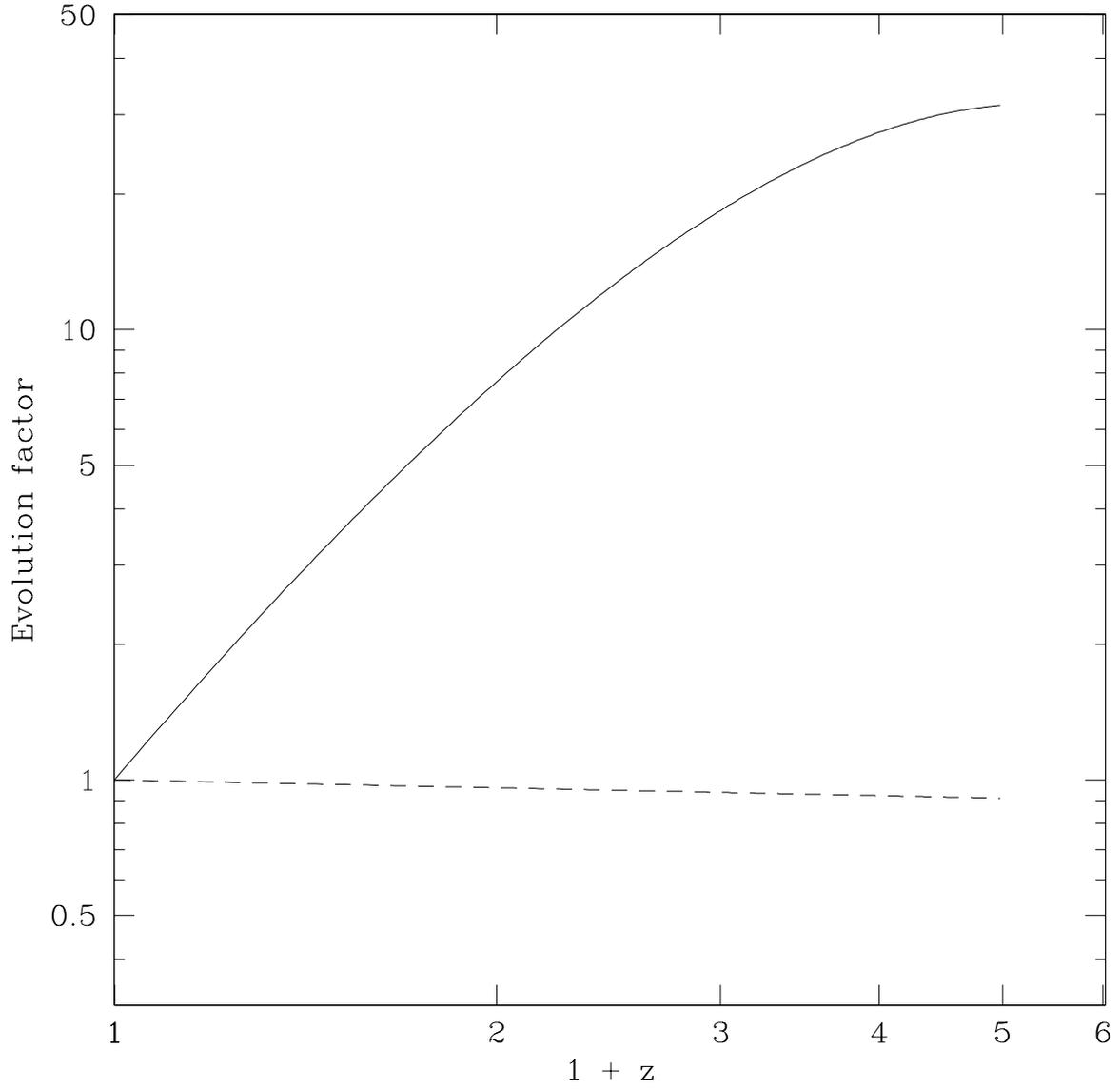}
\caption{Evolution functions for the fitted model found in 
\S\ref{lumfunc.modelfit}.  The solid line is $f(z)$
(luminosity evolution), and the dashed line is $g(z)$ (number density
evolution) (see eqs.~\ref{eq.lumevol} and \ref{eq.numevol}).  }
\label{fig.evolfactor}
\end{figure}

Our best fit is $\{ Q=3.97, q=1.02, z_q=1.39, P=-0.0579, p=23.1,
z_p=14.3 \}$.  The resulting evolution factors $f(z)$ and $g(z)$ are
plotted in Figure~\ref{fig.evolfactor}.  The $n(z)$ distribution
predicted by the model is shown in Figure~\ref{fig.nzlimit} (bottom
panel).  The peak of the model $n(z)$ distribution falls at a lower
redshift ($z\sim0.3$) than the peak in the data ($z\sim0.5$), but the
tail of the distribution is reasonable and the total number density
under the curve is similar (within 5\%) for the data and the model
(models with a peak at higher redshift tend to have a much shorter
tail or a larger total number of sources $n$ and thus a larger
discrepancy with the observed total).  The luminosity function
predicted by the model is shown as the curves in
Figure~\ref{fig.lumfunc_evolve}, and is a good fit to the data points,
except for the $z=0.05-0.2$ bin (which includes only two galaxies),
and the $z=0.6-0.9$ bin (where one point is too high and the other is
too low).  The model predicted star-forming radio background is
$3.0\times10^{-23}\wmsh$, which is a reasonable fit to the observed
value.  While the model does not perfectly match the three data
constraints, it is the best compromise between them.  Models that give
a better fit to the observed $n(z)$ {\em shape} result in a poor fit
to the other two data constraints.  For instance, some evolution
models can produce a longer tail on the $n(z)$ distribution, but that
raises the total background and the total surface density $n$
significantly above the observed levels.

In the early stages of this work, it seemed that the full 6 parameters
of our evolution model (eqs.~\ref{eq.lumevol}, \ref{eq.numevol}) were
necessary to achieve a good fit.  In the end, the best fit model shows
virtually no number density evolution, and only a mild turn-over in
luminosity evolution.  Pure luminosity evolution \{\ie\ $f(z)=(1+z)^Q$
and $g(z)=1$\} has often been suggested in the literature
\citep{rowan-robinson93a,hopkins98a}.  It turns out that pure
luminosity evolution with $Q=2.74$ gives a similar fit to the
luminosity function data points, but predicts a tail on the $n(z)$
distribution which extends beyond $z=5$, and a star-forming radio
background that is slightly too high ($4.1\times10^{-23}\wmsh$).

The 6 parameters used here are very interdependent.  As we progressed
through this work, adding data and model features, we fit our model to
the data numerous times.  The resulting fits occured in a wide range
of this parameter space, with the turnover at high redshift sometimes
occuring in $f(z)$ (luminosity evolution) and sometimes occuring in
$g(z)$ (number density evolution).  Some fits had a steeper increase
in $f(z)$ and a decrease in $g(z)$, others had a shallower increase in
$f(z)$ and an increase in $g(z)$.  This reminds us that the 6
parameters are degenerate, and most likely a different
parameterization with fewer free parameters could describe the data as
well.  Finding a new parameterization, however, would go beyond the
scope of this present work, given the limited sample size currently
available.  Although the shapes of $f(z)$ and $g(z)$ varied greatly
between different fits, all of the fits predicted generally similar
shapes for the observables (the $n(z)$ distribution and the luminosity
function), and for the predicted global star formation history (see
\S\ref{sfhist.model}).

\section{Star Formation History}
\label{sfhist}

Now that we have a model of the evolving luminosity function, we can
use it to determine the star formation history.  First we describe
the relationship between star formation rate and radio luminosity
(\S\ref{sfhist.radio}), then calculate the star formation history
directly from the data with minimal model dependence
(\S\ref{sfhist.data}), and finally calculate it from the model
(\S\ref{sfhist.model}).

\subsection{Star formation rate from radio luminosity}
\label{sfhist.radio}

For an individual star-forming galaxy, the star formation rate is
directly proportional to its radio luminosity \citep{condon92a}:
\begin{equation}
	\unit{SFR} = Q
	\left( \frac{L_\nu / (\whz)
		}{5.3\times10^{21} \left(\nu/\unit{GHz}\right)^{-0.8} 
	 	+ 5.5\times10^{20} \left(\nu/\unit{GHz}\right)^{-0.1} } 
	\right)\unit{\msun yr^{-1}}
\label{eq.sfrtoLpergal}
\end{equation}
\citet{condon92a} derives this relation by calculating the synchrotron
radio emission from supernova remnants (the first term in the
denominator) and the thermal radio emission from HII regions (the
second term).  The spectral index of 0.8 is typical for the
non-thermal component of a radio source at 1.4~GHz.  This relation is
derived purely from radio considerations.  \citet{cram98a}, however,
compares this relation to H$\alpha$ studies, and finds that they give
similar star formation rates for local individual galaxies, with the
except of galaxies with extremely large star formation rates.
(\citealp{hopkins00b} has found SFR-dependent dust corrections that
shift the optical results to match \ref{eq.sfrtoLpergal} and radio
observations.)

Both the thermal and non-thermal components of the radio expression
are proportional to the formation rate of high-mass stars ($M>5\msun$)
which produce supernova and large HII regions, so the factor $Q$ is
included to account for the mass of {\it all} stars in the interval
$0.1-100\msun$,
\begin{equation}
	Q = \frac{  \int_{0.1\msun}^{100\msun} M \psi(M) dM 
	        }{  \int_{  5\msun}^{100\msun} M \psi(M) dM 
                },		
\end{equation}
where $\psi(M)\propto M^{-x}$ is the initial mass function (IMF).  We
have assumed throughout a Salpeter IMF ($x=2.35$), for which $Q=5.5$.
If an upper limit of $125\msun$ is used, then $Q=5.4$.  If we use a
range of mass $0.25-100\msun$, as suggested by \citet*{gould96a} then
$Q=3.9$.  We will use $Q=5.5$ in the following.

Condon's relationship (eq.~\ref{eq.sfrtoLpergal}) uses the emitted
source luminosity at a frequency of 1.4~GHz, and thus the corrections
given in eq.~\ref{eq.LemitLband} must be applied.  We should also
consider whether there are other ways in which the connection between
SFR and radio luminosity might evolve with redshift.  At 1.4~GHz, the
thermal term in eq.~\ref{eq.sfrtoLpergal} is much smaller than the
synchrotron term, so evolution in the thermal term will have little
effect.  In the synchrotron term, the dependence of the emitted flux
on the supernova environment is weak \citep{condon92a}, so little
evolution is expected.  However, at high redshifts, relativistic
electrons may experience significant inverse-Compton cooling from the
intense FIR energy density or the cosmic microwave background.
Another effect that might cause significant evolution in
eq.~\ref{eq.sfrtoLpergal} is an evolving IMF, entering through the
factor $Q$.  In active starbursts, the IMF may be weighted to
high-mass stars \citep{elmegreen99a}, which would result in a smaller
value of $Q$.  However, the smallest $Q$ is unity (when virtually all
mass occurs in high-mass stars), so the strongest decrease in our
calculated star formation history from a radical change in the IMF
would be roughly a factor of five. Note that evolution of the IMF
would affect optical estimates of the star formation rate as well.  In
the following calculations we assume that eq.~\ref{eq.sfrtoLpergal}
does not evolve.

To determine the star formation rate per comoving volume, we simply
substitute the radio luminosity density (such as eq.~\ref{eq.lumdens}
or \ref{eq.lumdens.data}) for the source luminosity $L_\nu$ in
eq.~\ref{eq.sfrtoLpergal}, giving
\begin{equation}
  \Psi(z) = Q 
	  \left(\frac{\rho_{e,1.4}(z)}{ 4.6\times10^{21} \whzmpc}\right)
	  \myrmpc
\label{eq.sfhist}
\end{equation}
where 1.4\unit{GHz} is used in the denominator of
eq.~\ref{eq.sfrtoLpergal}, as all data have already been converted to
1.4 GHz in the rest frame.

\subsection{Star formation history estimated from the data}
\label{sfhist.data}

We now calculate the star formation history directly from the survey data in
\S\ref{data} by using the luminosity density of the detected sources.
For each redshift bin ($\zmin<z<\zmax$), the luminosity density is
\begin{equation}
	\rho_{e,1.4}(z) = \Sigma_i \frac{L_i B_i C(z)}{V_c(\zmin, \zmax(L_i))}
\label{eq.lumdens.data}
\end{equation}
where $B_i$ is the surface density given in eq.~\ref{eq.pbcor}, $C(z)$
is a correction for faint sources described below, $\zmax(L_i)$ is the
largest $z$ in the bin for which the luminosity of the source $L_i$
was above detection limit of survey, and $V_c$ is the comoving volume
given in eq.~\ref{eq.covol}.  This luminosity density can then be used
in eq.~\ref{eq.sfhist} to find the evolving star formation density.
Without the correction factor $C(z)$, the luminosity density includes
only individual sources brighter than the flux limit of the survey.
It does not include the luminosity density of sources too faint to be
detected individually, and so it clearly underestimates the star
formation rate (but calculations without $C(z)$ have the advantage of
being independent of our evolution model and provide a lower limit).

\begin{figure}
\epsscale{0.75}
\plotone{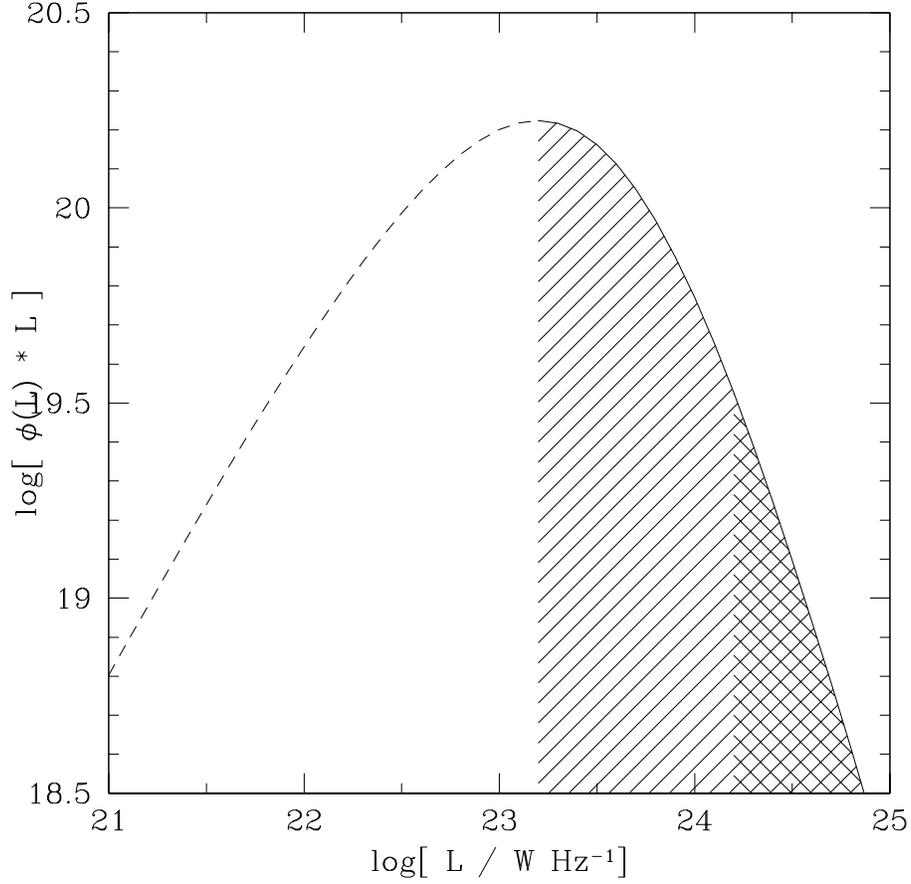}
\caption{ The integral under the $\phi(L) * L$ curve is proportional
to the luminosity density.  The relation for redshift $z=1.6$ is
shown.  The total luminosity density is due to all sources brighter
than $S_{8~GHz}\sim1\mujy$ {\it (hash)}, but only discrete sources
above $S_{8~GHz}\sim9\mujy$ are detected in the survey sample {\it
(cross-hash)}.  The ratio of the two regions gives the correction to
the luminosity density needed to account for sources too faint to be
detected individually in the survey; here it is about 3.8. }
\label{fig.lumdens_contr}
\end{figure}

To account for these faint sources, we use the evolving luminosity
function found in \S\ref{lumfunc.modelfit}.
Figure~\ref{fig.lumdens_contr} illustrates the faint source
correction, using a redshift of 1.6 as an example.  The integral under
the curve in the figure is proportional to the total luminosity
density.  The flux limit of the survey, however, only allows detection
of individual sources above a certain luminosity, \ie\ in the
cross-hash area.  The faint source correction factor $C(z)$ in
eq.~\ref{eq.lumdens.data} would then be the ratio of the total area to
the cross-hash area.  We have argued, however, that the slope of the
source number counts changes below about 1\mujy\ \citep{haarsma98a},
so that in fact most of the sources will occupy only the hashed area
of fig.~\ref{fig.lumdens_contr}.  Thus, a more realistic correction to
eq.~\ref{eq.lumdens.data} is the ratio of the hashed region to the
cross-hashed region, \ie\ the ratio of the luminosity density due to
sources brighter than 1\mujy\ to the luminosity density from sources
brighter than the flux limit of the survey,
\begin{equation}
	C(z) = \frac{ \int_{L(S_{lim},z)}^{\inf} \lel \; \phi(\lel,z) \dlogL 
	           }{ \int_{L(1\mujy ,z)}^{\inf} \lel \; \phi(\lel,z) \dlogL }
\label{eq.faintcorr}
\end{equation}
A list of corrections for several redshifts is given in
Table~\ref{tab.faintcorr}.  Note that if the slope of the number
counts of radio sources were assumed to stay the same below 1\mujy,
these corrections would be even larger, and so would the calculated star
formation density.

\epsscale{1}

\begin{figure}
\epsscale{0.75}
\plotone{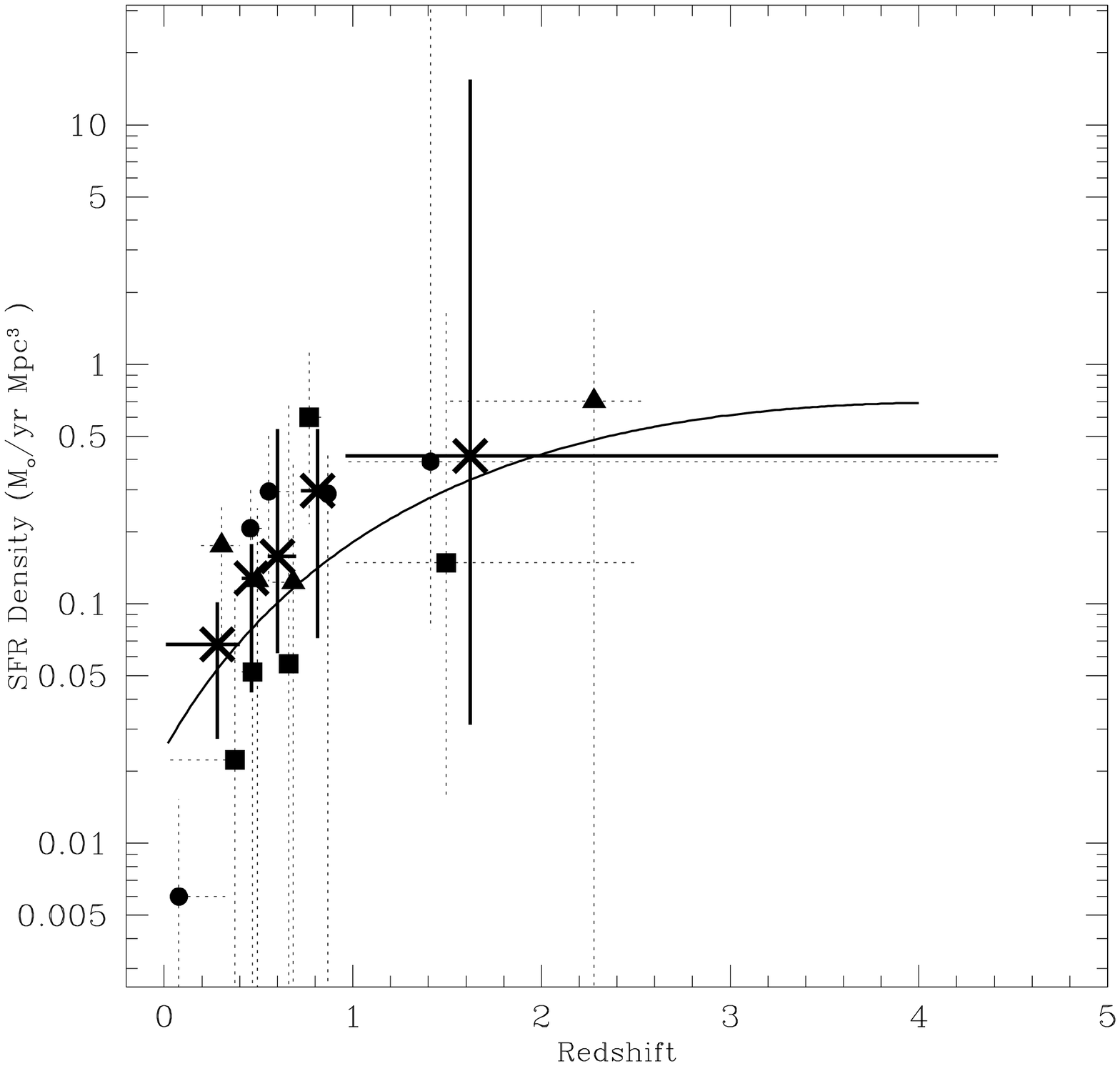}
\caption{Star formation history data points (see \S\ref{sfhist.data}).
Circles are from the HDF field, triangles from the SSA13 field, and
squares from the V15 field.  The average is shown with crosses and
solid error bars.  The curve is the star formation history predicted
by the model evolving luminosity function (note that this curve was
{\it not} fit to the data points shown here, see
\S\ref{sfhist.model}).  Vertical error bars are the larger of Poisson
errors or lower/upper limits (\S\ref{data.limits}).  Horizontal error
bars are the range of source redshifts in the ``upper'' sample.}
\label{fig.sfhist.radio}
\end{figure}

We calculated the star formation density using eqs.~\ref{eq.sfhist},
\ref{eq.lumdens.data}, \ref{eq.faintcorr} for the lower, middle, and
upper samples described in \S\ref{data.limits}; the results are shown
in Figures~\ref{fig.sfhist.radio} and \ref{fig.sfhist.all} and listed
in Table~\ref{tab.sfhist}.  Recall that the ``lower'' sample includes
only sources with spectroscopic redshifts and definite identifications
with spirals, irregulars, or mergers, and thus is the minimum amount
of star formation activity consistent with the data.  The ``middle''
value includes some sources with ambiguous identifications and rough
photometric redshifts, but is our best guess at the total
radio-selected star-forming population.  The ``upper'' sample includes
all the sources and is the maximum possible star formation activity
allowed by the radio data.  In Figure~\ref{fig.sfhist.radio}, the
points are values calculated from the middle sample, plotted at the
average of the source redshifts in the bin.  The vertical error bars
are either the limits from the lower and upper samples, or the Poisson
errors (from the number of galaxies per bin), whichever is larger (in
most cases, the lower limits and upper limits are larger than the
Poisson errors).  To make the lower sample a true lower limit, we did
{\it not include} the faint source correction $C(z)$.  Thus the lower
limit is for only those sources clearly identified with star forming
systems {\em and} having spectroscopic redshifts, with no allowance
made for evolution of the luminosity function or for sources below the
survey flux limits.  This is surely a gross underestimate of the true
star formation density value, since faint sources, unidentified
sources, and sources without spectroscopic redshifts are all missing,
but it provides a firm lower limit to the true star-formation rate.
The ``middle'' and ``upper'' samples do have the correction $C(z)$ for
faint sources applied.

\citet{mobasher99a} have done a similar calculation of star formation
density from a survey of faint radio sources.  They find no evidence
for evolution from $z=0$ to $z=1$, and a {\em decrease} in star
formation density above $z=0.3$.  Their results, however, are
based on a radio survey sample that is only 50\% complete, and thus
their results are highly dependent on assumptions made when correcting
for incompleteness.  Similarly, their optical identifications and
spectroscopic redshifts are much less complete than ours.  Finally,
the radio surveys we use extend to much fainter flux densities
where star formation is more likely to dominate the radio emission.
The fainter flux limit also allows us to detect more high redshift
sources.  Thus, we believe our results for the star formation density
are more reliable than those of \citet{mobasher99a}.

\subsection{Star formation history predicted by model }
\label{sfhist.model}

The star formation history can also be determined directly from the
evolution model found in \S\ref{lumfunc.modelfit}.  We simply
calculate the luminosity density emitted at 1.4~GHz
(eq.~\ref{eq.lumdens}), and use eq.~\ref{eq.sfhist} to find the star
formation density.  The resulting star formation density prediction is
the curve plotted in Figures~\ref{fig.sfhist.radio} and
\ref{fig.sfhist.all}.

The model curve in Figures~\ref{fig.sfhist.radio} and
\ref{fig.sfhist.all} falls somewhat below the averaged data points
(heavy crosses).  Note that the model curve was {\it not} fitted to
these averaged data points (which were calculated using
equations~\ref{eq.sfhist}, \ref{eq.lumdens.data}, \ref{eq.faintcorr}),
but rather the model was found from the evolving luminosity function
alone (using equations~\ref{eq.lumdens} and \ref{eq.sfhist}).  The
evolving luminosity function, in turn, was fitted to the $n(z)$ data,
the luminosity function data, and the radio background (see
\S\ref{lumfunc.modelfit}).  As discussed in \S\ref{lumfunc.modelfit},
our model fits these data well but not perfectly, so small differences
between the model prediction and the star formation data points are
not unreasonable.

This sort of calculation was previously done by \citet{cram98b}, using
the \citet{condon89a} luminosity function and \citet{condon84b}
evolution model.  Please note the typographical error in eq.~2 of
\citet{cram98b}, which differs by a factor of 28.2 from our
eq.~\ref{eq.sfrtoLpergal}; the correct values were used in their
calculations (L. Cram, private communication).  We agree with Cram's
calculation of 0.026\myrmpc\ for the local star formation density, and
calculation of star formation history from Condon's early model.

Note that these methods and results are much improved over our very
preliminary work \citep{haarsma99b}, which assumed that the majority
of detected faint radio sources lie at the redshift of peak star
formation activity.  In fact, the peak of the observed redshift
distribution (Figures~\ref{fig.nzztype}-\ref{fig.nzlimit}) is at a
lower redshift than the peak star formation activity
(Figures~\ref{fig.sfhist.radio} and \ref{fig.sfhist.all}), due to
cosmological factors, such as the dependence of the comoving volume on
redshift.

\section{Discussion and Conclusions}
\label{conclusion}

\begin{figure}
\plotone{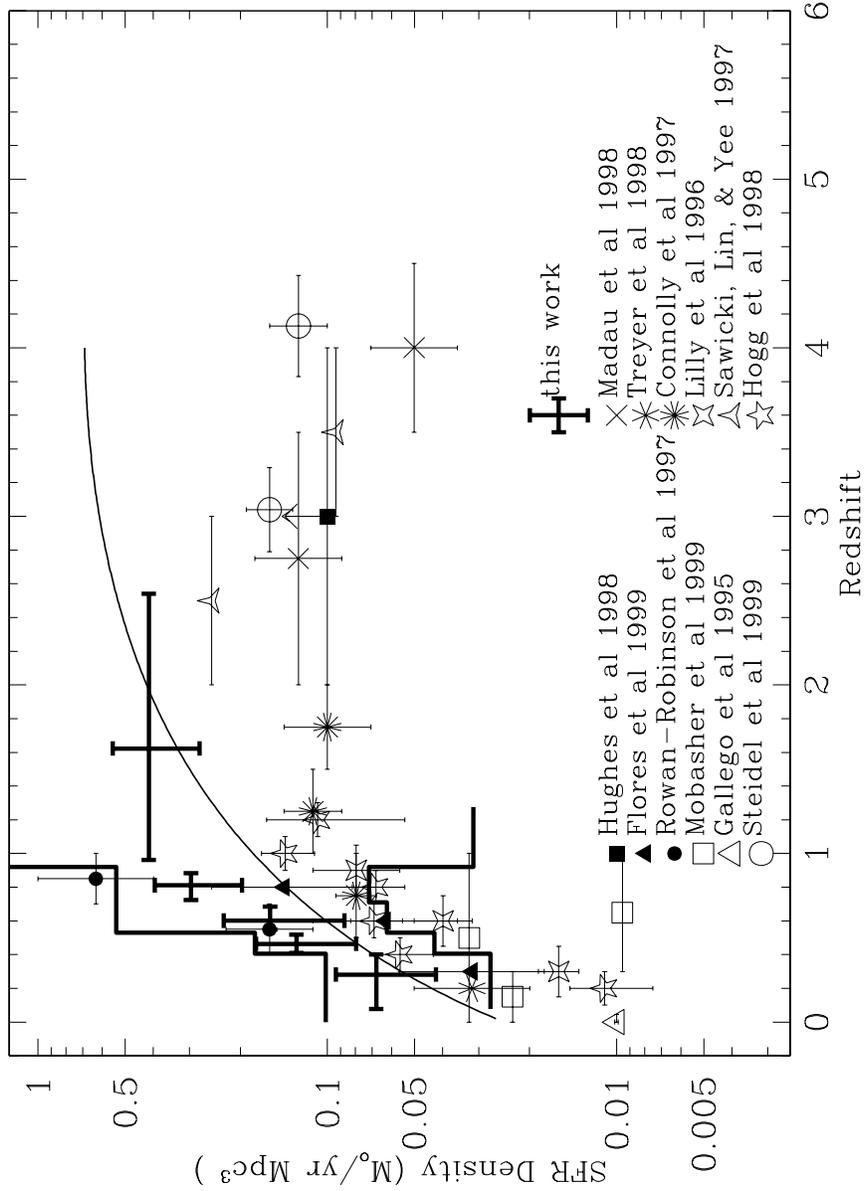}
\caption{ Star formation history. Data points are the same as in
Figure~\ref{fig.sfhist.all-radio}, with our results overlaid.  The
thick crosses show the star formation density of our middle sample
(defined in \S\ref{data.limits}), with error bars indicating Poisson
errors.  The thin curve is our model prediction, found in
\S\ref{sfhist.model} by fitting to the redshift distribution,
luminosity function, and extragalactic radio background (not to the
heavy crosses shown).  The thick lines indicate firm lower and upper
limits on the star formation density, calculated in
\S\ref{sfhist.data} using samples defined in \S\ref{data.limits}.  }
\label{fig.sfhist.all}
\end{figure}

Radio wavelength determinations of the Universal star formation
history have the important advantage of being independent of the dust
content of galaxies.  Additionally, it is possible to cull relatively
clean samples of star-forming objects using radio properties such as
spectral index, morphology, and variability.  Our results are shown in
Figure~\ref{fig.sfhist.all}, overlaid with the star formation
histories found in several other studies.  In
Figure~\ref{fig.sfhist.all}, the thick crosses show the star formation
density of our middle sample, with error bars indicating Poisson
errors.  Although it is possible that our middle sample may include
some low-luminosity AGN (Seyferts, etc.), our careful definition of
the sample (\S\ref{data.limits}) and the large fraction of sources
with clear optical identifications reduces this contamination.  The
thin curve is our model prediction, found in \S\ref{sfhist.model} by
fitting to the luminosity function, redshift distribution, and radio
background (not to the thick crosses).  The thick lines indicate our
lower and upper limits on the star formation density, which are
calculated in \S\ref{sfhist.data} using samples defined in
\S\ref{data.limits}.  The lower limit is very firm, since it includes
only those sources with spectroscopic redshifts and identifications
with spirals, irregulars and mergers, and does not include the star
formation in galaxies fainter than the survey detection limit.  The
upper limit is also firm, since it includes {\em all} detected radio
sources, even those not associated with star forming galaxies, but is
more uncertain than the lower limit since it includes sources without
spectroscopic redshifts.

If we were to assume a different cosmology, our results would change
somewhat.  The values of \ho, \Om, and \Ol\ affect the calculation of
distance from redshift, and luminosity from flux density.  The star
formation density is proportional to luminosity/volume, so it is
inversely proportional to distance and directly proportional to \ho.
If we had assumed \ho=100\kmsmpc\ instead of \ho=50\kmsmpc, our data
points and model for the star formation density would be twice as
large.  The values of \Om\ and \Ol\ also affect the distance
measurement.  As an extreme example, in a nearly empty but flat
universe (\Om=0.1, \Ol=0.9), the distance to a $z=1$ object is about
1.5 times larger than the distance to it in our assumed cosmology
(\Om=1, \Ol=0), and thus the star formation density would be about 2/3
of our listed value.  Note that all data points in
Figures~\ref{fig.sfhist.all-radio} and \ref{fig.sfhist.all} depend
similarly on these cosmological parameters.

At low redshifts, we agree with the findings of many studies that star
formation density increases rapidly from the local universe to $z=1$.
We disagree with \citet*{cowie99b} who find a gradual (rather than
steep) increase from $z=0$ to $z=1$, and with \citet{mobasher99a} who
find a {\em decrease} in star formation density $z=0$ to $z=1$.  Our
firm lower limit is significantly higher $z<1$ than the {\it
extinction corrected} optical results of \citet{lilly96a} at
$z\lesssim0.7$, indicating that some star formation has been obscured
by dust.  Our ``middle sample'' data points fall above all the optical
and ultraviolet studies shown, indicating that these studies have
probably missed some star formation by underestimating the dust
extinction (see \citealp{hopkins00b} for SFR-dependent dust
corrections that bring these data more into agreement with our radio
results).  Our results are similar to the star formation density from
the Infrared Space Observatory (ISO) survey of the HDF
\citep{rowan-robinson97a}.
 
At redshifts above $z=1$, we cannot draw strong conclusions.  There
are few sources with spectroscopic redshifts in this range, so our
calculations are based in large part on less secure photometric
redshift estimates and random redshift assignments for the very red
objects (\S\ref{data.z}).  Our assumption that the relationship
between radio luminosity and star formation rate does not evolve also
becomes less sure as we move to higher redshift (see discussion in
\S\ref{sfhist.radio}).  The Initial Mass Function may also be
evolving, although this would affect optical and ultraviolet estimates
of star formation history as well.  Finally, the faint source
corrections (eq.~\ref{eq.faintcorr}, Table~\ref{tab.faintcorr}) become
larger at high redshift, and thus depend more strongly on the assumed
shape of the luminosity function.  In fact, at redshifts above 1.5,
the current radio survey limits only probe the extreme end of the
luminosity function (SFR per galaxy $> 1000\msun$).  Deeper surveys
are needed to detect radio counterparts to typical high-redshift
optical objects, \eg\ Lyman break galaxies (for instance, the
predicted radio fluxes for even the most luminous Lyman break galaxies
in the HDF are only a few \mujy\ at 1.4~GHz, \eg\
\citealp*{meurer99a}).  Planned improvements to the VLA will allow
future surveys to reach this sensitivity.
 
Still, our calculations at high redshift show that if even a small
number of star-forming radio sources exist beyond $z\sim1.5$, they
would indicate a large, optically-hidden fraction of star-formation
density.  In particular, the population of radio sources with faint,
red optical counterparts may be dust-enshrouded
(\citealp{richards99d,barger00a,waddington99a}, \S\ref{data.id}), and
hence missed even in the deepest optical and ultraviolet studies.
Deeper high resolution radio observations, accompanied by close to
complete spectroscopic identifications, are needed to determine
accurately the amount of ``hidden'' star formation in the early
Universe.

\nocite{hughes98a}
\nocite{flores99a}
\nocite{rowan-robinson97a}
\nocite{tresse98a}
\nocite{gallego95a}
\nocite{steidel99a}
\nocite{madau98a}
\nocite{treyer98a}
\nocite{connolly97a}
\nocite{lilly96a}
\nocite{hogg98a}
\nocite{baugh98a}

\section*{Acknowledgments}

D.B.H. and R.B.P. acknowledge the support of NSF grant AST 96-16971
and the Keck Northeast Astronomy Consortium. D.B.H. acknowledges the
support of a Cottrell College Science Award from Research Corporation.
E.A.R. acknowledges support from Hubble Fellowship Grant
HG-0123.01-99A, and from STScI which is operated by AURA under NASA
contract NAS 5-26555.  R.A.W. acknowledges support from AST-8821016
and AST-9802963 from the National Science Foundation and NASA grants
AR-6948.04-96A and GO-7452.01-96A from STScI under NASA contract
NAS5-26555.  We are grateful to James Lowenthal for providing us with
spectroscopic redshifts in advance of publication.  We thank Ian
Waddington, Ed Fomalont, Ken Kellermann, and Andrew Hopkins for
helpful comments and discussions, and our anonymous referee for
detailed and helpful comments on the manuscript.

\newpage
\bibliography{apj-jour,radio}
\bibliographystyle{apj}

\newpage  

\begin{deluxetable}{lllccrrrr}
\tabletypesize{\scriptsize}
\tablecaption{Summary of Radio Surveys\label{tab.data}}
\tablehead{
\colhead{Field} & \colhead{Location} & \colhead{Band} 
& \colhead{Flux limit at field center} & \colhead{Field size (amin$^2$)}
& \colhead{$N$} & \colhead{$N_{sp}$} & \colhead{$N_{ph}$} & \colhead{$N_a$}
}

\startdata
Hubble Deep Field  & 12h+62d  & 8 GHz & 9\mujy     & 66          & 29  & 19       & 4        & 6   \\
\citep{richards98a}&          &       &            &             &     &          &          &     \\[6pt] 
SSA13 field        & 13h+42d  & 8 GHz & 8.8\mujy   & \phn7       & 15  & 8        & 3        & 4   \\
\multicolumn{3}{l}{\citep{windhorst95a}} &         &             &     &          &          &     \\[6pt]	
V15 field          & 14h+52d  & 5 GHz & 16\mujy    & 86          & 33  & 18       & 3        & 12  \\ 
\multicolumn{3}{l}{\citep{fomalont91a,hammer95a}} & &            &     &          &          &     \\[6pt]
Total              &          &       &            &             & 77  & 45       & 10       & 22  \\	
\enddata
\tablecomments{The listed flux limit for each field is approximately 5
times the RMS noise at the beam center.  The field size is the region
in which both radio and optical data are available (see
Tables\ref{tab.hdf}, \ref{tab.13h}, and \ref{tab.v15} for details).
$N$ is the total number of sources above the flux limit, $N_{sp}$ is
the number of those sources with spectroscopic redshifts, $N_{ph}$ is
the number with redshifts estimated from I or HK$'$ band magnitudes,
and $N_{a}$ is the number with redshifts randomly assigned (see
\S\ref{data.z} for how assignments were made). }

\end{deluxetable}


\begin{deluxetable}{lrrrcllllrlcrl}
\tabletypesize{\tiny}
\tablecolumns{14}

\tablehead{
\colhead{Name} & \colhead{$S_8$} & \colhead{$S_{1.4}$} & \colhead{$\alpha$} & \colhead{$B_i$} &
\colhead{galtype} & \colhead{ztype} & \colhead{$z$} & \colhead{$HK'$} &
\colhead{$I$} & \colhead{$z_{max}$} & \colhead{$\log(\lel)$} & \colhead{SFR} & \colhead{Sample}  
\\
              & \colhead{(\mujy)} & \colhead{(\mujy)} &                     & \colhead{(1/asec$^2$)} &
                  &                 &               & \colhead{(mag)} &
\colhead{(mag)}&                     & \colhead{($\log[W/Hz]$)} & \colhead{($\msun$/yr)} &                   
\\
\colhead{1} & \colhead{2} & \colhead{3} & \colhead{4} & \colhead{5} &
\colhead{6} & \colhead{7} & \colhead{8} & \colhead{9} & \colhead{10} &
\colhead{11} & \colhead{12} & \colhead{13} & \colhead{14}
}
\tablecaption{HDF\label{tab.hdf}}

\startdata
   3632+1105 & 21.8        & (23) & $<$0.03 & 0.037 & sim     & sp\tablenotemark{d} & 0.518 & 17.22      & 19.79        & 0.785 & 22.49    & 37      & UML    
\\ 3634+1212 & 56.5        & 211  & 0.74    & 0.019 & sim     & sp\tablenotemark{cd}& 0.458 & 16.38      & 19.10        & 0.966 & 23.45    & 340     & UML    
\\ 3634+1240 & 52.6        & 198  & 0.74    & 0.020 & sim     & sp\tablenotemark{ac}& 1.219 & 19.11      & 22.29        & 2.34  & 24.49    & 3700    & UML    
\\ 3637+1135 & 17.5        & (23) & $<$0.15 & 0.049 & sim     & sp\tablenotemark{a} & 0.078 & 17.07      & 18.61        & 0.108 & 20.78    & 1       & UML    
\\ 3640+1010 & 29.2        & 65   & 0.45    & 0.029 & fr      & afr                 & 1.1   & 21.22      &$>$25\phd\phn & 1.78  & 23.80    & 750     & U      
\\ 3641+1142 & 18.6        & 30   & 0.27    & 0.045 & sim     & sp\tablenotemark{d} & 0.548 & 19.26      & 22.07        & 0.759 & 22.70    & 61      & UML    
\\ 3642+1331 & 79.9        & 432  & 0.94    & 0.016 & el      & sp\tablenotemark{e} & 4.42  & 21.23      &$>$25\phd\phn & 9.12  & 26.52    & \nodata & U      
\\ 3642+1545 & 53.6        & 131  & 0.50    & 0.020 & un(sim) & sp\tablenotemark{cd}& 0.857 & 18.07      & 20.86        & 1.76  & 23.85    & 860     & UM     
\\ 3644+1133 & 752\phd\phn & 1290 & 0.30    & 0.015 & el      & sp\tablenotemark{ac}& 1.050 & 17.64      & 21.04        & 6.38  & 25.00    & \nodata & U      
\\ 3644+1249 & 10.2        & (21) & (0.4)   & 0.249 & sim     & sp\tablenotemark{a} & 0.557 & 18.30      & 20.59        & 0.589 & 22.59    & 47      & UML    
\\ 3646+1404 & 190\phd\phn & 177  & $-$0.04 & 0.015 & sim     & sp\tablenotemark{abc}&0.962 & 18.19      & 20.75        & 3.85  & 23.95    & 1100    & UML    
\\ 3646+1447 & 13.3        & 77   & 0.98    & 0.081 & el      & ph                  & 0.69  & 20.21      & 22.05        & 0.804 & 23.50    & $<$390  & U      
\\ 3646+1448 & 24.7        & 112  & 0.84    & 0.033 & fr      & afr                 & 1.5   &$>$22\phd\phn &$>$25\phd\phn & 2.16 & 24.52   & 4000    & U      
\\ 3649+1313 & 14.0        & 51   & 0.72    & 0.072 & sim     & sp\tablenotemark{ac}& 0.475 & 18.68      & 21.10        & 0.575 & 22.87    & 90      & UML    
\\ 3651+1030 & 26.0        & 99   & 0.74    & 0.031 & sim     & sp\tablenotemark{abc}&0.410 & 17.37      & 19.89        & 0.638 & 23.01    & 120     & UML    
\\ 3651+1221 & 16.8        & 60   & 0.71    & 0.052 & fr      & afr                 & 1.7   &$>$21.5\phn &$>$25\phd\phn & 2.16  & 24.34    & 2700    & U      
\\ 3652+1444 & 185\phd\phn & 148  & $-$0.12 & 0.015 & el      & sp\tablenotemark{abc}&0.322 & 16.53      & 18.70        & 1.35  & 22.84    & $<$83   & U      
\\ 3653+1139 & 15.1        & 60   & 0.77    & 0.062 & sim     & sp\tablenotemark{cd}& 1.275 & 19.44      & 21.90        & 1.55  & 24.04    & 1300    & UML    
\\ 3655+1311 & 12.3        & (23) & $<$0.35 & 0.102 & el      & sp\tablenotemark{d} & 0.968 & 18.62      & 21.62        & 1.11  & 23.18    & $<$180  & U      
\\ 3657+1455 & 15.3        & (23) & $<$0.23 & 0.061 & un(sim) & sp\tablenotemark{d} & 0.859 & 18.83      & 21.44        & 1.07  & 23.01    & 120     & UM 
\\ 3700+0908 & 66.7        & 326  & 0.89    & 0.018 & fr      & afr                 & 2.9   & 20.98      &$>$25\phd\phn & 5.75  & 25.82    & \nodata & U  
\\ 3701+1146 & 29.5        & 98   & 0.67    & 0.028 & fr      & sp\tablenotemark{cd}& 0.884 & 20.14      & 24.81        & 1.41  & 23.80    & 770     & UM 
\\ 3707+1408 & 29.0        & 49   & 0.29    & 0.029 & fr      & ph                  & 2.07  & 20.29      & 24.70        & 3.35  & 24.28    & 2300    & UM 
\\ 3708+1056 & 26.4        & 49   & 0.35    & 0.031 & sim     & sp\tablenotemark{cd}& 0.423 & 17.91      & 19.87        & 0.684 & 22.68    & 57      & UML
\\ 3708+1246 & 19.5        & (23) &$<$0.09  & 0.042 & sim     & ph                  & 0.887 & 19.82      & 21.73        & 1.24  & 23.01    & 120     & UM 
\\ 3711+1331 & 31.1        & 108  & 0.69    & 0.027 & sim     & ph                  & 1.53  & 19.59      & 22.75        & 2.42  & 24.47    & 3600    & UM 
\\ 3716+1512 & 85.8        & 180  & 0.41    & 0.016 & sim     & sp\tablenotemark{c} & 0.558 & 17.36      & 19.80        & 1.45  & 23.53    & 410     & UML
\\ 3721+1129 & 630\phd\phn & 383  &$-$0.28  & 0.015 & fr      & afr                 & 2.5   & 20.90      &$>$25\phd\phn & $>$10 & 25.07    & \nodata & U  
\\ 3725+1128 & 530\phd\phn & 5960 & 1.35    & 0.015 & fr      & afr                 & 2.7   &$>$21.5\phn &$>$25\phd\phn & 9.12  & 27.25    & \nodata & U      

\enddata
\tablenotetext{a}{\citet{cohen96a}}
\tablenotetext{b}{\citet{lowenthal97a}}
\tablenotetext{c}{\citet{barger00a}}
\tablenotetext{d}{\citet{cohen00b}}
\tablenotetext{e}{\citet{waddington99a}}

\tablecomments{ These sources are cataloged by \citet{richards98a}
who gives the detailed source positions.  Sources are included here
only if they are within 4.6$'$ of the radio field center and above
the 8~GHz radio flux limit of \citet{richards98a} (additional sources
were detected in \citealp{richards00a} but are not included because of
their complicated primary beam corrections and lack of spectroscopic
redshifts).  The flux measurements, $S_8$ and $S_{1.4}$, are from
\citet{richards00a}.  Several sources were identified as faint/red by
\citet{richards99d}.  The I and HK$'$ magnitudes are from
\citet{barger99a}.  The remaining quantities in the table are
calculated or assigned in this work.  }
\end{deluxetable}


\begin{deluxetable}{lrrrclllrrlcrl}
\tabletypesize{\tiny}
\tablecolumns{14}

\tablehead{
\colhead{Name} & \colhead{$S_8$} & \colhead{$S_{1.4}$} & \colhead{$\alpha$} & \colhead{$B_i$} &
\colhead{galtype} & \colhead{ztype} & \colhead{$z$} & \colhead{$HK'$} &
\colhead{$I$} & \colhead{$z_{max}$} & \colhead{$\log(\lel)$} & \colhead{SFR} & \colhead{Sample}  
\\
              & \colhead{(\mujy)} & \colhead{(\mujy)} &                     & \colhead{(1/asec$^2$)} &
                  &                 &               & \colhead{(mag)} &
\colhead{(mag)}&                     & \colhead{($\log[W/Hz]$)} & \colhead{($\msun$/yr)} &                   
\\
\colhead{1} & \colhead{2} & \colhead{3} & \colhead{4} & \colhead{5} &
\colhead{6} & \colhead{7} & \colhead{8} & \colhead{9} & \colhead{10} &
\colhead{11} & \colhead{12} & \colhead{13} & \colhead{14}
}

\tablecaption{SSA13\label{tab.13h}}

\startdata

   (\phn1) 1214+3822 & 8.9     & (18) & (0.4) & 0.245 & fr      & afr & 1.5   & 21.95  &$>25$\phd\phn& 1.51  & 23.56 & 440     & U    
\\ (\phn3) 1215+3703 & 14.1    & (29) & (0.4) & 0.141 & un(sim) & sp  & 0.322 & 18.70  & 20.81       & 0.398 & 22.19 & 19      & UM   
\\ (\phn5) 1216+3921 & 33.7    & (69) & (0.4) & 0.141 & fr      & afr & 2.3   & $>25$  &$>25$\phd\phn& 3.94  & 24.60 & 4800    & U    
\\ (\phn6) 1217+3913 & 24.1    & (49) & (0.4) & 0.141 & un(sim) & sp  & 0.494 & 17.57  & 19.99       & 0.767 & 22.84 & 84      & UM   
\\ (\phn7) 1218+3931 & 15.1    & (31) & (0.4) & 0.141 & fr      & ph  & 2.54  & 21.22  &$>25$\phd\phn& 3.16  & 24.36 & 2800    & UM   
\\ (\phn9) 1218+3844 & 26.6    & (45) & (0.4) & 0.141 & un(el)  & sp  & 0.698 & 17.57  & 20.62       & 1.12  & 23.23 & $<$210  & U    
\\    (10) 1220+3833 & 18.3    & (37) & (0.4) & 0.141 & un(sim) & a   & 0.559 & 21.78  & 24.08       & 0.767 & 22.85 & 85      & U    
\\    (11) 1220+3703 & 14.3    & (29) & (0.4) & 0.141 & fr      & ph  & 2.02  & 20.77  &$>25$\phd\phn& 2.45  & 24.08 & 1500    & UM   
\\    (12) 1220+3923 & 14.3    & (29) & (0.4) & 0.141 & fr      & afr & 1.9   & 23.36  &$>25$\phd\phn& 2.32  & 24.02 & 1300    & U    
\\    (13) 1221+3923 & 10.1    & (21) & (0.4) & 0.225 & sim     & sp  & 0.302 & 17.08  & 19.42       & 0.323 & 21.99 & 12      & UML  
\\    (15) 1221+3723 & 18.3    & (37) & (0.4) & 0.141 & sim     & sp  & 0.180 & 17.04  & 18.99       & 0.254 & 21.76 & 7       & UML  
\\    (16) 1222+3827 & 17.4    & (36) & (0.4) & 0.141 & un(sim) & ph  & 0.685 & 19.38  & 22.03       & 0.923 & 23.03 & 130     & UM   
\\    (18) 1223+3909 & 21.1    & (43) & (0.4) & 0.141 & un(el)  & sp  & 0.765 & 17.19  & 19.92       & 1.11  & 23.23 & $<$200  & U    
\\    (19) 1224+3712 & 29.2    & (60) & (0.4) & 0.141 & un(sim) & sp  & 0.401 & 16.68  & 19.14       & 0.684 & 22.72 & 63      & UM   
\\    (24) 1228+3800 & 23.8    & (49) & (0.4) & 0.141 & sim     & sp  & 0.316 & 17.04  & 19.60       & 0.495 & 22.40 & 30      & UML  
\\

\enddata

\tablecomments{ These sources are cataloged by \citet{windhorst95a}
(catalog number in parentheses) who gives the detailed source
positions; the source name is derived from its radio position.
Sources are included here if they are above the radio flux limit of
\citet{kellermann00a}, fully within the optical field of
\citet{windhorst95a}, and not identified with quasars or stars.  
The I and HK$'$ magnitudes are from \citet{cowie96a}.  The $S_8$ flux
densities are from \citet{kellermann00a}.  Sources 1, 5, 7, 11, and 12
are classified as faint/red by \citet{richards99d}.  The spectroscopic
redshifts for sources 3, 13, 15, and 19 are from \citet{windhorst94c},
and for sources 6, 9, 18, and 24 are from J. Lowenthal (private
communication).  The remaining quantities in the table are calculated
or assigned in this work.  
}

\end{deluxetable}

\begin{deluxetable}{lrrrcllllrlcrl}
\tabletypesize{\tiny}
\tablecolumns{14}

\tablehead{
\colhead{Name} & \colhead{$S_5$} & \colhead{$S_{1.4}$} & \colhead{$\alpha$} & \colhead{$B_i$} &
\colhead{galtype} & \colhead{ztype} & \colhead{$z$} & \colhead{$HK'$} &
\colhead{$I$} & \colhead{$z_{max}$} & \colhead{$\log(\lel)$} & \colhead{SFR} & \colhead{Sample}  
\\
              & \colhead{(\mujy)} & \colhead{(\mujy)} &                     & \colhead{(1/asec$^2$)} &
                  &                 &               & \colhead{(mag)} &
\colhead{(mag)}&                     & \colhead{($\log[W/Hz]$)} & \colhead{($\msun$/yr)} &                   
\\
\colhead{1} & \colhead{2} & \colhead{3} & \colhead{4} & \colhead{5} &
\colhead{6} & \colhead{7} & \colhead{8} & \colhead{9} & \colhead{10} &
\colhead{11} & \colhead{12} & \colhead{13} & \colhead{14}
}

\tablecaption{V15\label{tab.v15}}

\startdata
   15V05   & 41    & 96     & 0.68  & 0.019 & sim    & sp  & 0.989 &\nodata & 22.04        & 1.42   & 23.92 & 1000    & UML    
\\ 15V10   & 1912  & 2560   & 0.23  & 0.012 & el     & a   & 0.180 & 20.86  & 22.27        & 1.53   & 23.58 & $<$460  & U      
\\ 15V11   & 70    & 171    & 0.71  & 0.014 & sim    & sp  & 0.375 &\nodata & 19.86        & 0.700  & 23.16 & 170     & UML    
\\ 15V15   & 24    & (40)   & (0.4) & 0.034 & el     & a   & 0.302 &\nodata & 22.30        & 0.363  & 22.27 & $<$22   & U      
\\ 15V18   & 44    & (73)   & (0.4) & 0.018 & fr     & ph  & 2.0   & 21.1   &$>$25\phd\phn & 2.99   & 24.47 & 3600    & UM     
\\ 15V19   & 24    & (43)   & (0.4) & 0.034 & sim    & sp  & 0.754 &\nodata & 21.17        & 0.902  & 23.17 & 180     & UML    
\\ 15V21   & 298   & 807    & 0.80  & 0.012 & sim    & sp  & 0.724 &\nodata & 21.29        & 2.19   & 24.53 & 4100    & UML    
\\ 15V23   & 54    & (80)   &$<$0.31& 0.016 & el*    & sp  & 1.149 &\nodata & 21.08        & 1.91   & 23.89 & 930     & U
\\ 15V24   & 79    & 141    & 0.46  & 0.014 & sim    & sp  & 0.660 & 19.22  & 20.64        & 1.29   & 23.60 & 480     & UML
\\ 15V26a  & 23    & (38)   & (0.4) & 0.037 & el*    & sp  & 0.372 &\nodata & 21.87        & 0.442  & 22.45 & $<$34   & U  
\\ 15V26b  & 24    & (40)   & (0.4) & 0.034 & el*    & sp  & 0.155 &\nodata & 19.27        & 0.188  & 21.64 & $<$5    & U  
\\ 15V28   & 74    & (80)   &$<$0.6 & 0.014 & el     & sp  & 0.988 & 19.99  & 21.74        & 1.95   & 23.66 & $<$550  & U  
\\ 15V33   & 20    & (33)   & (0.4) & 0.054 & el     & a   & 0.410 &\nodata & 23.2\phn     & 0.457  & 22.49 & $<$37   & U  
\\ 15V34   & 1311  & 1160   &$-$0.10& 0.012 & el     & sp  & 0.838 &\nodata & 22.01        & 6.46   & 24.62 & \nodata & U  
\\ 15V37   & 51    & (80)   &$<$0.36& 0.016 & el     & ph  & 2.52  & 20.53  & 23.65        & 4.03   & 24.74 & \nodata & U  
\\ 15V39   & 35    & (58)   & (0.4) & 0.021 & el     & sp  & 0.992 & 19.89  & 21.85        & 1.38   & 23.62 & $<$500  & U  
\\ 15V40   & 33    & 106    & 0.93  & 0.022 & el     & sp  & 0.976 & 19.67  & 21.94        & 1.29   & 24.02 & $<$1300 & U  
\\ 15V45   & 33    & 150    & 1.21  & 0.022 & el     & a   & 0.475 &\nodata & 24.3\phn     & 0.631  & 23.42 & $<$320  & U  
\\ 15V47   & 53    & 131    & 0.72  & 0.016 & sim    & sp  & 0.809 & 22.23  & 23.00        & 1.30   & 23.85 & 850     & UML  
\\ 15V48   & 20    & (33)   & (0.4) & 0.054 & sim    & sp  & 0.743 &\nodata & 21.11        & 0.822  & 23.08 & 150     & UML
\\ 15V49   & 31    & 96     & 0.90  & 0.023 & un(sim)& ph  & 0.537 & 19.22  & 19.96        & 0.700  & 23.31 & 180     & UM 
\\ 15V50   & 705   & 1722   & 0.71  & 0.012 & un(el) & a   & 0.654 &\nodata & 22.76        & 2.79   & 24.74 & \nodata & U  
\\ 15V51   & 41    & (68)   & (0.4) & 0.019 & un(sim)& a   & 0.724 &\nodata & 23.14        & 1.08   & 23.36 & 280     & U  
\\ 15V53   & 30    & 85     & 0.83  & 0.024 & sim    & a   & 0.754 &\nodata & 23.04        & 0.966  & 23.61 & 490     & U  
\\ 15V57   & 37    & (61)   & (0.4) & 0.020 & el*    & sp  & 0.010 & 17.83  & 17.70        & 0.0153 & 19.40 & $<1$    & U  
\\ 15V59   & 19    & (31)   & (0.4) & 0.066 & un(sim)& a   & 0.960 &\nodata & 24.5\phn     & 1.04   & 23.32 & 250     & U  
\\ 15V60   & 39    & (64)   & (0.4) & 0.019 & sim    & sp  & 0.812 &\nodata & 21.94        & 1.19   & 23.46 & 350     & UML
\\ 15V62   & 24    & (40)   & (0.4) & 0.034 & fr     & afr & 2.1   &$>$21.3 &$>$25\phd\phn & 2.48   & 24.26 & 2200    & U  
\\ 15V67   & 46    & (76)   & (0.4) & 0.017 & fr     & afr & 2.5   &\nodata &$>$25\phd\phn & 3.80   & 24.73 & \nodata & U  
\\ 15V70   & 576   & 2414   & 1.14  & 0.012 & fr     & afr & 1.3   &\nodata &$>$25\phd\phn & 4.27   & 25.80 & \nodata & U  
\\ 15V72   & 46    & (76)   & (0.4) & 0.017 & el     & a   & 1.219 & 20.9   & 23.8\phn     & 1.88   & 23.96 & 1100    & U  
\\ 15V73   & 37    & (61)   & (0.4) & 0.020 & el     & sp  & 0.746 &\nodata & 20.95        & 1.07   & 23.35 & $<$270  & U  
\\ 15V81   & 28    & (46)   & (0.4) & 0.026 & el*    & sp  & 1.158 &\nodata & 22.16        & 1.46   & 23.69 & $<$590  & U  
\\
\enddata

\tablecomments{ These sources are cataloged by \citet{fomalont91a},
who gives detailed source positions and the 5~GHz and 1.4~GHz flux
densities.  Sources are not included here if they are quasars, stars,
below the radio flux limit, or outside the optical field of
\citet{hammer95a}.  Sources 30, 36, 41, and 69 were found later to be
pairs of radio sources, with both members of the pair below the radio
flux limit, and so are not included.  The galaxy types, spectroscopic
redshifts, and I and K magnitudes are from \citet{hammer95a}.  An
asterisk (*) indicates that \citet{hammer95a} classified the galaxy
type based on its emission lines, which they find to be more
characteristic of AGN than starbursts.  The spectroscopic redshifts
for 15V5 and 15V47, and the galaxy type for 15V47, are from
\citet{brinchmann98a}.  \citet{lilly99a} finds a photometric redshift
of 2 for 15V18.  \citet{flores99a} identifies source 23 as a Seyfert 2
galaxy.  The remaining quantities in the table are calculated or
assigned in this work.  }

\end{deluxetable}

\begin{deluxetable}{ccc}
\tablecolumns{3}

\tablecaption{Faint Source Correction Factors\label{tab.faintcorr}}

\tablehead{
Redshift        & \multicolumn{2}{c}{Correction factor $C(z)$} \\ 
                & 8 GHz, $S_{lim}=9\mujy$  & 5 GHz, $S_{lim}=16\mujy$ \\
}
\startdata
0.28		& \phn1.3		& \phn1.5	\\
0.46 		& \phn1.6		& \phn2.0	\\
0.60		& \phn1.9		& \phn2.3	\\
0.81		& \phn2.2		& \phn3.0	\\
1.6\phn	        & \phn3.8		& \phn5.8	\\
\enddata

\end{deluxetable}

\begin{deluxetable}{cc|ccc}
\tablecolumns{5}
\tabletypesize{\small}
\tablecaption{Star Formation History\label{tab.sfhist}}

\tablehead{
\multicolumn{2}{c}{Redshift}        & \multicolumn{3}{c}{SFR Density (\myrmpc)} \\ 
\colhead{Average\tablenotemark{a}} & \colhead{Range\tablenotemark{b}} &
\colhead{Value from Middle Sample} & \colhead{Poisson error range} & \colhead{Lower and Upper Limits}
}
\startdata
0.28     & 0.010-0.401   & 0.068 & 0.042-0.093    & 0.027-0.101 \\
0.46     & 0.410-0.518   & 0.128 & 0.080-0.176    & 0.043-0.178 \\
0.60     & 0.548-0.698   & 0.158 & 0.087-0.228    & 0.062-0.537 \\
0.81     & 0.724-0.884   & 0.296 & 0.197-0.395    & 0.072-0.536\\
1.6\phn  & 0.960-4.42\phn& 0.414 & 0.276-0.552    & 0.031-15.5\phn \\
\enddata
\tablenotetext{a}{average redshift in bin from middle sample}
\tablenotetext{b}{range of redshifts in bin from upper sample}

\end{deluxetable}


\end{document}